\documentclass[12pt]{article}

\usepackage{dsfont}
\usepackage{epsfig}
\usepackage{amsfonts}
\usepackage{amssymb}
\usepackage{amsmath}
\usepackage{bbm}
\usepackage{cite}

\newcommand{\bmat}{\left(\begin{array}}
\newcommand{\emat}{\end{array}\right)}

\def\gtrsim{\mathrel{\raise.3ex\hbox{$>$\kern-.75em\lower1ex\hbox{$\sim$}}}}

\def\ds{\displaystyle}

\def\a{\alpha}

\def\b{\beta}
\def\g{\gamma}
\def\d{\delta}

\def\-{\hphantom{-}}

\def\s2{\frac{1}{\sqrt2}}

\def\wt{\widetilde}

\def\beq{\begin{equation}}
\def\eeq{\end{equation}}
\def\beqa{\begin{eqnarray}}
\def\eeqa{\end{eqnarray}}
\def\D{{\rm D}}

\def\im{{\rm Im \,}}
\def\re{{\rm Re \,}}

\def\Tr{{\rm Tr \,}}

\def\T{{\rm T}}
\def\M{{\rm M}}
\def\Z{{\mathbb Z}}
\def\RR{{\mathbb R}}
\def\diag{{\rm diag \,}}

\def\eps{\epsilon}

\def\cc{{\mathcal C}}
\def\cd{{\mathcal D}}

\def\cj{{\mathcal J}}
\def\ck{{\mathcal K}}

\def\eps{{\epsilon }}
\def\cw{{\mathcal W}} 

\def\cam{{\mathcal M}}
\def\cn{{\mathcal N}}
\def\cp{{\mathcal P}}
\def\ct{{\mathcal T}}
\def\cv{{\mathcal V}}

\def\cz{{\mathcal Z}}

\def\cs{{\mathcal S}}
\def\cx{{\mathcal X}}

\def\mg{m_{3/2}}
\def\mg2{m^2_{3/2}}

\def\deq#1{\mbox{$D$=#1}}
\def\neq#1{\mbox{$\cn$=#1}}
\def\Dsl{\,\raise.15ex\hbox{/}\mkern-13.5mu D} 

\newcommand{\msm}[1]{\mbox{\small$#1$}}
\newcommand{\mfn}[1]{\mbox{\footnotesize$#1$}}
\begin{document}
\pagestyle{plain}

\makeatletter
\@addtoreset{equation}{section}
\makeatother
\renewcommand{\theequation}{\thesection.\arabic{equation}}
\pagestyle{empty}
\rightline{AEI-2008-078, \  IFT-UAM/CSIC-08-52}
\vspace{5mm}
\begin{center}
\LARGE{\bf Algebras and  non-geometric flux vacua
\\[5mm]}
\large{
Anamar\'{\i}a Font${}^{a,c}$, Adolfo Guarino${}^b$ and Jes\'us M. Moreno${}^b$
 \\[3mm]}
\small{
${}^a$
Departamento de F\'{\i}sica, Centro de F\'{\i}sica Te\'orica y Computacional, \\[-0.3em]
Facultad de Ciencias, Universidad Central de Venezuela, \\[-0.3em]
A.P. 20513, Caracas 1020-A, Venezuela. \\[2mm]
${}^b$
Instituto de F\'{\i}sica Te\'orica UAM/CSIC,\\[-0.3em]
Facultad de Ciencias C-XVI, Universidad Aut\'onoma de Madrid, \\[-0.3em]
Cantoblanco, 28049 Madrid, Spain\\[2mm]
${}^c$
Max-Planck-Institut f\"ur Gravitationsphysik, Albert-Einstein-Institut\\[-0.3em]
14476 Golm, Germany
\\[1cm]}
\small{\bf Abstract} \\[3mm]
\end{center}
{\small
In this work we classify the subalgebras satisfied by non-geometric $Q$-fluxes in type IIB orientifolds on $\msm{\T^6/(\Z_2 \times \Z_2)}$ with 
three moduli $(S,T,U)$. We find that there are five subalgebras compatible with the symmetries, each one leading to a characteristic
flux-induced superpotential. Working in the 4-dimensional effective supergravity we obtain families of supersymmetric ${\rm AdS}_4$ vacua with 
all moduli stabilized at small string coupling $g_s$. Our results are mostly analytic thanks to a judicious parametrization of the non-geometric, 
RR and NSNS fluxes. We are also able to leave the flux-induced $C_4$ and $C_8$ RR tadpoles as free variables, thereby enabling us to
study which values are allowed for each $Q$-subalgebra. Another novel outcome is the appearance of multiple vacua for special sets of fluxes. 
However, they generically have $g_s > 1$ unless the net number of O3/D3 or O7/D7 sources needed to cancel the tadpoles is large. 
We also discuss briefly the issues of axionic shift symmetries and cancellation of Freed-Witten anomalies. 
}


\newpage
\setcounter{page}{1}
\pagestyle{plain}
\renewcommand{\thefootnote}{\arabic{footnote}}
\setcounter{footnote}{0}
\tableofcontents

\newpage

\section{Introduction}
\label{sec:intro}

The study of flux compactifications in string theory has been pursued intensively in recent years \cite{generalreviews}.
One important motivation is the possibility to stabilize the massless moduli at a minimum of the potential induced by the fluxes.
The simplest scenarios for this mechanism are provided by type IIB and type IIA \neq1 orientifolds with $p$-form fluxes turned on \cite{generalreviews}.
In IIA compactifications the mixture of NSNS and RR fluxes generates a superpotential that depends on all closed string
moduli allowing to stabilize them without invoking non-perturbative effects \cite{Grimm, Derendinger, vz1, DeWolfe, cfi}.
Moreover, in the IIA setup it is natural to add the so-called geometric $f$-fluxes that determine the isometry algebra of 
the internal space  \cite{Derendinger, vz1, cfi}. The case of nilpotent algebras was studied in \cite{gmpt, andriot, caviezel} and an example with 
internal $\mathfrak{su(2)^2}$ was spelled out in \cite{af}.  

To recover T-duality between IIA and IIB compactifications, it is necessary to introduce new parameters referred to as
non-geometric fluxes \cite{stw1, stw2, wecht}. 
The original observation is that performing a T-duality to NSNS $\bar H$-fluxes leads to geometric $f$-fluxes \cite{glmw, kstt}.
Further T-dualities give rise to generalized $Q$ and $R$-fluxes \cite{stw1}. The $Q$'s are called non-geometric because the emerging
background after two T-dualities can be described locally but not globally. The third T-duality is formal, evidence for the $R$-fluxes
comes rather from T-duality at the level of the effective superpotential \cite{stw1}. Moreover, the $Q$ and $R$-fluxes logically
extend \cite{stw1, dabholkar} the set of structure constants of the gauge algebra, generated by isometries and shifts of the $B$ field, 
that is known to contain the geometric and NSNS fluxes \cite{ss, km}.

In this article we consider type IIB orientifolds with O3/O7-planes in which only NSNS $\bar H$ and non-geometric $Q$-fluxes are invariant
under the orientifold action. These fluxes together induce a superpotential that depends on all closed string moduli.
One advantage of working with IIB is that the $Q$-fluxes by themselves appear as the structure constants
of a subalgebra of the full gauge algebra. However, one must keep in mind that the $\bar H$ and $Q$ in IIB map into all kinds of
fluxes in type IIA with O6-planes, and into non-geometric $R$ plus geometric $f$ in IIB with O9/O5-planes. Similar examples
with generalized fluxes have been considered by several authors \cite{stw1, acfi, stw2, vz2, benmachiche, tasinato,ihl, palti, camara}.  

Our guiding principle is precisely the classification of the subalgebras satisfied by the non-geometric $Q$-fluxes. 
We will discuss a simplified scheme with additional symmetries in order to reduce the number of fluxes. 
Concretely, we study compactification on $\msm{(\T^2 \times \T^2 \times \T^2)/(\Z_2 \times \Z_2)}$,
and further impose invariance under exchange of the internal $\T^2$'s. In this way we obtain the same model with moduli $(S,T,U)$ 
proposed in \cite{stw1} and generalized in \cite{acfi}. We have classified the allowed subalgebras of the $Q$-fluxes of
the $(S,T,U)$-model. There are five inequivalent classes, namely $\mathfrak{so(4)}$,  $\mathfrak{so(3,1)}$, $\mathfrak{su(2)+u(1)^3}$,
$\mathfrak{iso(3)}$ and the nilpotent algebra denoted $n(3.5)$ in \cite{gmpt}. The non-semisimple solutions are contractions of
$\mathfrak{so(4)}$ consistent with the symmetries. A compelling byproduct is that each subalgebra yields a characteristic flux-induced
superpotential. The corresponding 12-dimensional gauge algebras can be easily identified after a convenient change of basis.   
 
We are mostly interested in discovering supersymmetric flux backgrounds with non-geometric fluxes switched on,
and all moduli stabilized. To this end we work exclusively with the \deq4 effective action.
We widen the search of vacua of \cite{stw2} in several respects. A key difference is that in most cases  
we can solve the F-flat conditions analytically and can therefore derive explicit expressions for the moduli vevs in terms
of the fluxes. The computations are facilitated by using a transformed complex structure \mbox{$\cz=\msm{(\a U + \b)/(\g U + \d)}$}, 
invariant under the modular group $SL(2,\Z)_U$. The independent non-geometric fluxes are precisely parametrized by 
$\Gamma=\genfrac{(}{)}{0pt}{}{\a \, \b}{\g \, \d}$. The parametrization of NSNS and RR fluxes is also dictated by $\Gamma$.
By exploiting the variable $\cz$ we can effectively factor out the vacuum degeneracy due to modular transformations. 

There is a further vacuum degeneracy originating from special constant translations in the axions $\re S$ and $\re T$. We argue that vacua connected
by this type of translations are identical because the full background including the RR fluxes is invariant under such axionic shifts. 

In our analysis the values of the flux-induced $C_4$ and $C_8$ RR tadpoles are treated as variables. 
To cancel these tadpoles in general requires to add D-branes besides the orientifold planes. These D-branes are also constrained
by cancellation of Freed-Witten anomalies \cite{cfi, vz2}.
In our concrete setup, D3-branes and unmagnetized D7-branes wrapping an internal $\T^4$ are free of anomalies and can be
included. However, such D-branes do not give rise to charged chiral matter. 

By treating the flux tadpoles as variables we can deduce in particular that the vacua 
found in \cite{stw2}, having O3-planes and no O7/D7 sources, can only arise when the $Q$-subalgebra is the compact $\mathfrak{so(4)}$.  
For completeness we study the supersymmetric ${\rm AdS}_4$ minima due to the fluxes of all compatible $Q$-subalgebras, including the
non-compact $\mathfrak{so(3,1)}$. In general, such vacua exist in all cases but unusual types of sources might be needed to cancel
the tadpoles. Interestingly, in models based on semisimple subalgebras we find that there can exist more than one vacuum for some 
combinations of fluxes. 

It is well known that supersymmetric or no-scale Minkowski vacua in IIB orientifolds with RR and NSNS fluxes require sources of
negative RR charge such as O3-planes or wrapped D7-branes \cite{gkp}. However, working with the effective \deq4 formalism we find that O3-planes 
and/or D7-branes can be bypassed in fully stabilized supersymmetric ${\rm AdS}_4$ vacua, provided specific non-geometric fluxes are turned on.
It is conceivable that such vacua only occur in the effective theory and will not survive after lifting to a full string background.
Helpful hints in this direction can come from our results relating properties of the vacua with the gauge algebra.
It might well be that only models built on certain algebras can be lifted to full backgrounds. 
The newly proposed formulation of non-geometric fluxes based on compactification on doubled twisted tori suggests that the gauge algebra has to be
compact or admit a discrete cocompact subgroup \cite{reid, prezas}. It is also feasible that the 
recent description of non-geometric fluxes in the context of generalized geometry \cite{minasian} could be applied to deduce the generalized flux
configurations which allow supersymmetric vacua. A discussion of these issues is beyond our present scope.

We now outline the paper. In section \ref{sec:gen} we review the properties of the fluxes and write down the flux-induced effective
quantities needed to investigate the vacua. The classification of the $Q$-subalgebras is carried out in section \ref{sec:alg}, where we also obtain the
parametrization of the non-geometric and NSNS fluxes that is crucial in the subsequent analysis. In section \ref{sec:newvars} we introduce
the transformed complex structure $\cz$ motivated by modular invariance. Using this variable then points to the efficient parametrization of the
RR fluxes given in the appendix. In the end we are able to derive very compact expressions for the flux-induced superpotential and tadpoles according to 
the particular $Q$-subalgebra. In section \ref{sec:vac} we solve the F-flat conditions and collect the results that distinguish the vacua with moduli stabilized.
The salient features of these vacua are discussed in section \ref{sec:lands}. Section \ref{sec:end} is devoted to some final comments.

\section{Generalities}
\label{sec:gen}

In this section we outline our notation to describe the non-geometric fluxes introduced in \cite{stw1}. To be specific we will work in 
the context of toroidal orientifolds with O3/O7-planes. We will discuss the case of generic untwisted moduli, and also the simpler 
isotropic model considered in \cite{stw1}.

\subsection{Fluxes}
\label{ssec:fluxes}

The starting point is a type IIB string compactification on a six-torus $\T^6$ whose basis of
1-forms is denoted $\eta^a$. Moreover, we assume the factorized geometry
\beq
\T^6=\T^2 \times  \T^2 \times \T^2  \,\,:\,\, 
(\eta^{1}\,,\,\eta^{2})  \,\,\times\,\,( \eta^{3}\,,\,\eta^{4}  )
\,\,\times\,\,( \eta^{5}\,,\,\eta^{6} )  \ .
\label{factorus}
\eeq
As in \cite{stw1}, we will use greek indices $\alpha,\beta,\gamma$ for horizontal 
$\,``-"$ $x$-like directions $(\eta^{1},\eta^{3},\eta^{5})$ and latin indices $i,j,k$ 
for vertical $\,``|"$ $y$-like directions $(\eta^{2},\eta^{4},\eta^{6})$ in the 2-tori.

The $\Z_2$ orientifold involution denoted $\sigma$ acts as
\beq
\sigma \ : \  ( \eta^{1}\,,\,\eta^{2}\,,\,\eta^{3}\,,\,\eta^{4}\,,\,\eta^{5}\,,\,\eta^{6} ) 
\ \rightarrow \ 
( -\eta^{1}\,,\,-\eta^{2}\,,\,-\eta^{3}\,,\,-\eta^{4}\,,\,-\eta^{5}\,,\,-\eta^{6} ) \ .
\label{osigma}
\eeq
There are 64 O3-planes located at the fixed points of $\sigma$.
We further impose a $\Z_2 \times \Z_2$ orbifold symmetry with generators acting as
\beqa
\label{orbifold1}
\theta_1 & : & ( \eta^{1}\,,\,\eta^{2}\,,\,\eta^{3}\,,\,\eta^{4}\,,\,\eta^{5}\,,\,\eta^{6} ) 
\ \rightarrow \ 
(\eta^{1}\,,\,\eta^{2}\,,\,-\eta^{3}\,,\,-\eta^{4}\,,\,-\eta^{5}\,,\,-\eta^{6} )  \ , \\[2mm]
\theta_2 & : & (\eta^{1}\,,\,\eta^{2}\,,\,\eta^{3}\,,\,\eta^{4}\,,\,\eta^{5}\,,\,\eta^{6} ) 
\ \rightarrow \ (-\eta^{1}\,,\,-\eta^{2}\,,\,\eta^{3}\,,\,\eta^{4}\,,\,-\eta^{5}\,,\,-\eta^{6} )
 \ .
\nonumber
\eeqa
Clearly, there is another order-two element $\theta_3 = \theta_1 \theta_2$.
Under this $\Z_{2} \times \Z_{2}$ orbifold group, only 3-forms with one leg in each 2-torus survive. 
This also occurs in the compactification with an extra $\Z_{3}$ cyclic permutation of the three 2-tori 
that was studied in \cite{stw1, stw2}. In that case there are only O3-planes and two geometric
moduli, namely the overall K\"ahler and complex structure parameters.
In contrast, in our setup, the full symmetry group $\Z_2^3$ includes additional orientifold actions
$\sigma \theta_I$ that have fixed 4-tori and lead to \mbox{O$7_I$-planes}, $I=1,2,3$. 
Another difference is that in principle we have one K\"ahler and one complex structure parameter for each 
2-torus $\T_I^2$. 

The K\"ahler form and the holomorphic 3-form that encode the geometric moduli of the internal space
can be written in a basis of invariant forms that also enters in the description
of background fluxes. Under the $\Z_2 \times \Z_2$ orbifold action the invariant 3-forms are just     
\beq
\begin{array}{lclclcl}
\alpha_{0}=\eta^{135} & \quad ; \quad &  \alpha_{1}=\eta^{235} & \quad ; \quad & 
\alpha_{2}=\eta^{451} & \quad ; \quad & \alpha_{3}=\eta^{613}   \ , \\
\beta^{0}=\eta^{246} & \quad ; \quad & \beta^{1}=\eta^{146} & \quad ; \quad & 
\beta^{2}=\eta^{362} & \quad ; \quad & \beta^{3}=\eta^{524}  \ .
\end{array}
\label{basisab}
\eeq
where, e.g.  $\eta^{135}= \eta^1 \wedge \eta^3 \wedge \eta^5$. 
Clearly, these forms are all odd under the orientifold involution $\sigma$.
On the other hand, the invariant 2-forms and their dual 4-forms are
\beq
\begin{array}{lclcl}
\omega_{1}=\eta^{12} & \quad ; \quad & \omega_{2}=\eta^{34} & \quad ; \quad &
\omega_{3}=\eta^{56}  \ , \\
\tilde{\omega}^{1}=\eta^{3456} & \quad ; \quad & \tilde{\omega}^{2}=\eta^{1256} & \quad ; \quad & 
\tilde{\omega}^{3}=\eta^{1234}  \ .
\end{array}
\label{inv2form}
\eeq
These forms are even under $\sigma$. 
We choose the orientation and normalization
\beq
\int_{\M_6} \!\! \eta^{123456}=\cv_6 \ .
\label{normal1}
\eeq
The positive constant $\cv_6$ gives the volume of the internal space that we generically denote $\M_6$.
Notice that the basis satisfies
\beq 
\int_{\M_6} \!\! \alpha_{0} \wedge  \beta^{0}= -\cv_6 \quad , \quad
\int_{\M_6} \!\! \alpha_{I} \wedge  \beta^{J}= 
\int_{\M_6}\!\! \omega_{I}  \wedge \tilde{\omega}^{J}= \cv_6 \delta_{I}^{J} \quad \, , \quad \, I,J=1,2,3.
\label{normal2}
\eeq
The $\Z_{2} \times \Z_{2}$ orbifold symmetry restricts the period matrix $\tau^{ij}$ to be diagonal.
Then, up to normalization, the holomorphic 3-form is given by
\beq
\label{holoexpan}
\Omega= (\eta^1 + \tau_1 \eta^2) \wedge (\eta^3 + \tau_2 \eta^4) \wedge (\eta^5 + \tau_3 \eta^6)
=\alpha_{0} + \tau_{K}	\,\alpha_{K} + \beta^{K}\,\frac{\tau_1 \tau_{2}  \tau_{3}}{ \tau_{K}} + 
\beta^{0}\,\tau_1  \tau_{2}  \tau_{3} \ ,
\eeq
with the $H^{3}(\M_6,\Z)$ basis displayed in (\ref{basisab}).

The next step is to switch on background fluxes for the NSNS and RR 3-forms. Since both $H_3$ and $F_3$ are
odd under the orientifold involution, the allowed background fluxes can be expanded as
\beqa
\bar H_3 & = & b_{3} \,\alpha_{0} + b_2^{(I)} \,\alpha_{I}  + b_{1}^{(I)} \,\beta^{I} + b_{0} \,\beta^{0} \ , 
\label{H3expan} \\[2mm]
\bar F_3 & = & a_{3} \,\alpha_{0} + a_{2}^{(I)}  \,\alpha_{I}  + 
a_{1}^{(I)} \,\beta^{I} + a_{0} \,\beta^{0} \ .
\label{F3expan} 
\eeqa
All flux coefficients are integers because the integrals
of  $\bar H_3$ and $\bar F_3$ over 3-cycles are quantized. To avoid subtleties with exotic orientifold planes
we take all fluxes to be even \cite{frey, kst}.

As argued originally in \cite{glmw, kstt}, applying one T-duality transformation to the NSNS fluxes
can give rise to geometric fluxes $f^a_{bc}$ that correspond to structure constants of the isometry algebra
of the internal space. Performing further T-dualities leads to generalized fluxes denoted $Q_c^{ab}$ and $R^{abc}$ \cite{stw1}. 
The  $Q_c^{ab}$ are called non-geometric fluxes because the resulting metric after two T-dualities yields a background that is 
locally but not globally geometric \cite{stw2, wecht}. Compactifications with $R^{abc}$ fluxes are not even
locally geometric but these fluxes are necessary to maintain T-duality between type IIA and type IIB. 
The geometric and the R-fluxes must be even under the orientifold involution and are thus totally absent in type IIB with O3/O7-planes. 
On the other hand, the non-geometric fluxes must be odd and are fully permitted.

The main motivation of this work is to study supersymmetric vacua in toroidal type IIB orientifolds
with NSNS, RR and non-geometric $Q$-fluxes turned on. In our construction, the $\Z_2 \times \Z_2$
symmetry only allows 24 components of the flux tensor $Q_c^{ab}$, namely those with one leg on each
2-torus. This set of non-geometric fluxes is displayed in table \ref{tableNonGeometric}. All components
of the tensor $Q$ are integers that we take to be even. 
\begin{table}[htb]
\begin{center}\begin{tabular}{|c|c|c|}
\hline
Type & Components & Fluxes \\
\hline
\hline
$Q_{-}^{--} \equiv Q_{\alpha}^{\beta \gamma}$ & $ Q_{1}^{35}\,,\,Q_{3}^{51}\,,\,Q_{5}^{13}$ & 
$\tilde{c}_{1}^{\,(1)}\,,\,\tilde{c}_{1}^{\,(2)}\,,\,\tilde{c}_{1}^{\,(3)}$ \\
\hline
\hline
$Q_{|}^{|-} \equiv Q_{k}^{i \beta} $ & $ Q_{4}^{61}\,,\,Q_{6}^{23}\,,\,Q_{2}^{45}$ & 
$ \hat{c}_{1}^{\,(1)}\,,\,\hat{c}_{1}^{\,(2)}\,,\,\hat{c}_{1}^{\,(3)}$ \\
\hline
\hline
$Q_{|}^{-|} \equiv Q_{k}^{\alpha j}$ & $ Q_{6}^{14}\,,\,Q_{2}^{36}\,,\,Q_{4}^{52}$ & 
$ \check{c}_{1}^{\,(1)}\,,\,\check{c}_{1}^{\,(2)}\,,\,\check{c}_{1}^{\,(3)}$ \\
\hline
\hline
$Q_{|}^{--} \equiv Q_{k}^{\alpha\beta}$ & $Q_{2}^{35}\,,\,Q_{4}^{51}\,,\,Q_{6}^{13}$ &  
$ c_{0}^{\,(1)}\,,\,c_{0}^{\,(2)}\,,\,c_{0}^{\,(3)}$\\
\hline
\hline
$Q_{-}^{||} \equiv Q_{\gamma}^{i j}$ & $ Q_{1}^{46}\,,\,Q_{3}^{62}\,,\,Q_{5}^{24}$ & 
$ c_{3}^{\,(1)}\,,\,c_{3}^{\,(2)}\,,\,c_{3}^{\,(3)}$ \\
\hline
\hline
$Q_{-}^{|-} \equiv Q_{\gamma}^{i \beta}$ & $Q_{5}^{23}\,,\,Q_{1}^{45}\,,\,Q_{3}^{61}$ & 
$\check{c}_{2}^{\,(1)}\,,\,\check{c}_{2}^{\,(2)}\,,\,\check{c}_{2}^{\,(3)}$ \\
\hline
\hline
$Q_{-}^{-|} \equiv Q_{\beta}^{\gamma i}$ & $ Q_{3}^{52}\,,\,Q_{5}^{14}\,,\,Q_{1}^{36}$ & 
$\hat{c}_{2}^{\,(1)}\,,\,\hat{c}_{2}^{\,(2)}\,,\,\hat{c}_{2}^{\,(3)}$ \\
\hline
\hline
$Q_{|}^{||} \equiv Q_{k}^{i j}$ & $Q_{2}^{46}\,,\,Q_{4}^{62}\,,\,Q_{6}^{24}$ & 
$\tilde{c}_{2}^{\,(1)}\,,\,\tilde{c}_{2}^{\,(2)}\,,\,\tilde{c}_{2}^{\,(3)}$ \\
\hline
\end{tabular}\end{center}
\caption{Non-geometric $Q$-fluxes.}
\label{tableNonGeometric}
\end{table}

\subsection{Effective action}
\label{ssec:action}

The NSNS, RR and non-geometric fluxes induce a potential for the closed string moduli.
We will focus on the untwisted moduli of the toroidal orientifold.
To write explicitly the effective action, recall first that 
the axiodilaton and the complex structure moduli are given by
\beq
S = C_0 + i e^{-\phi} \qquad ; \qquad U_I = \tau_I \quad ; \quad I=1,2,3 \ ,
\label{sumoduli}
\eeq 
where $C_0$ is the RR 0-form, $\phi$ is the 10-dimensional dilaton and the $\tau_I$ are
the components of the period matrix. The K\"ahler moduli $T_I$ are instead extracted from the expansion
of the complexified K\"ahler 4-form $\cj$, i.e. $\cj=-\sum T_{I} \, \tilde{\omega}^{I}$. In turn,
the real (axionic) part of $\cj$ arises from the RR 4-form $C_4$ whereas the imaginary part is
$e^{-\phi} J\wedge J/2$, where $J$ is the fundamental K\"ahler form. In fact, $\im T_I$ is basically
the area of the 4-cycle dual to the 4-form $\tilde\omega^I$. 

We are interested in compactifications that preserve \neq1 supersymmetry in four dimensions.
In this case we know that the scalar potential can be computed from the K\"ahler potential and
the superpotential.
The K\"ahler potential for the moduli is given by the usual expression
\beq
K =-\sum_{K=1}^{3}\log\left( -i\,(U_{K}-\bar{U}_{K})\right)  - \,\log\left( -i\,(S-\bar{S})\right)  -
\sum_{K=1}^{3} \log\left( -i\,(T_{K}-\bar{T}_{K})\right) \ , 
\eeq
which is valid to first order in the string and sigma model perturbative expansions.
The NSNS and RR fluxes induce a superpotential only for $S$ and the $U_I$. 
In absence of non-geometric fluxes K\"ahler moduli do not enter 
in the superpotential and non-perturbative effects such as gaugino condensation are required to get  
vacua with all moduli fixed. The $Q$-fluxes generate new couplings involving K\"ahler fields, thereby
opening the possibility to stabilize all types of closed string moduli.

The general superpotential can be computed from \cite{acfi}
\beq
\label{WInt}
W=\int_{\M_{6}} \!\!\! \left(G_{3} \,+\,Q\,\mathcal{J} \right) \,\wedge\, \Omega  \ ,
\eeq
where $G_{3}= \bar F_{3}-\,S\,\bar H_{3}$, and $Q\cj$ is a 3-form with components defined by 
\beq
(Q\,\mathcal{J})_{abc}=\frac{1}{2} \,Q_{[a}^{mn}\, \mathcal{J}_{bc]\,mn}  \ .
\label{qjcomp}
\eeq
Being a 3-form, $Q\cj$ can be expanded in the basis (\ref{basisab}). We obtain
\beq
\label{QJexpan}
Q\,\mathcal{J}=T_{K} \left( c_{3}^{(K)} \,\alpha_{0} - \cc_{2}^{(I K)} \,\alpha_{I}  - \cc_{1}^{(I K)} \,\beta^{I} 
+ c_{0}^{(K)} \,\beta^{0}  \right) \ ,
\eeq
where $\cc_1$ and $\cc_2$ are the non-geometric flux matrices
\beq
\cc_{1}=\left(
\begin{array}{lll}
-\tilde{c}_{1}^{\,(1)} & \check{c}_{1}^{\,(3)}   & \hat{c}_{1}^{\,(2)}    \\
 \hat{c}_{1}^{\,(3)}   & -\tilde{c}_{1}^{\,(2)}  & \check{c}_{1}^{\,(1)}  \\
 \check{c}_{1}^{\,(2)} & \hat{c}_{1}^{\,(1)}     & -\tilde{c}_{1}^{\,(3)} \\
\end{array}
\right)
\qquad ,\qquad
\cc_{2}=\left(
\begin{array}{lll}
-\tilde{c}_{2}^{\,(1)} & \check{c}_{2}^{\,(3)}   & \hat{c}_{2}^{\,(2)}    \\
 \hat{c}_{2}^{\,(3)}   & -\tilde{c}_{2}^{\,(2)}  & \check{c}_{2}^{\,(1)}  \\
 \check{c}_{2}^{\,(2)} & \hat{c}_{2}^{\,(1)}     & -\tilde{c}_{2}^{\,(3)} \\
\end{array}
\right) \ .
\label{c1c2mat}
\eeq
The expansion for the 3-form $G_3$ that combines the NSNS and the RR fluxes can be read off from (\ref{H3expan}) and (\ref{F3expan}).
Substituting the expansions of the holomorphic 3-form and the background fluxes in (\ref{WInt}) shows that the superpotential takes the form
\beq
W=P_{1}(U) + P_{2}(U)\,S + \sum_{K=1}^{3} P_{3}^{\,(K)}(U)\,T_{K} \ .
\label{fullW}
\eeq
The $P$'s are cubic polynomials in the complex structure moduli given by
\beqa
P_{1}(U) & = & a_{0} -\sum_{K=1}^{3} a_{1}^{\,(K)}\,U_{K} + 
\sum_{K=1}^{3} a_{2}^{\,(K)} \frac{U_{1}U_{2}U_{3}}{U_{K}} - a_{3} U_{1}U_{2}U_{3}   \ ,
\label{p1gen} \\[2mm]
P_{2}(U) & = & -b_{0} +\sum_{K=1}^{3} b_{1}^{\,(K)}\,U_{K} - 
\sum_{K=1}^{3} b_{2}^{\,(K)} \frac{U_{1}U_{2}U_{3}}{U_{K}} + b_{3} U_{1}U_{2}U_{3}  \ ,  
\label{p2gen} \\[2mm]
P_{3}^{\,(K)}(U) & = & c_{0}^{\,(K)} +\sum_{L=1}^{3} \cc_{1}^{\,(L K)}\,U_{L} - 
\sum_{L=1}^{3} \cc_{2}^{\,(L K)} \frac{U_{1}U_{2}U_{3}}{U_{L}} -c_{3}^{\,(K)} U_{1}U_{2}U_{3}  \ .
\label{p3gen}
\eeqa
The main feature of the flux superpotential is that it depends on all untwisted closed string moduli.

At this point we have a model with seven moduli whose potential depends on forty flux
parameters. Finding vacua in this generic setup is rather cumbersome. For this reason we consider
a simpler configuration in which the fluxes are isotropic. Concretely, we make the Ansatz 
\beqa
\tilde{c}_{1}^{\,(I)} \equiv \tilde{c}_{1}  \quad  ; \!\! &{}&
\hat{c}_{1}^{\,(I)} \equiv \hat{c}_{1} \quad ; \quad  
\check{c}_{1}^{\,(I)} \equiv \check{c}_{1}  \quad ; \quad 
\tilde{c}_{2}^{\,(I)} \equiv \tilde{c}_{2}  \quad ; \quad
\hat{c}_{2}^{\,(I)} \equiv \hat{c}_{2}  \quad  ; \quad
\check{c}_{2}^{\,(I)} \equiv \check{c}_{2}  \ ,
\nonumber \\
&{}& b_{1}^{\,(I)}\equiv b_{1}   \quad ; \quad 
b_{2}^{\,(I)} \equiv b_{2}  \quad ; \quad
a_{1}^{\,(I)} \equiv a_{1}   \quad ; \quad 
a_{2}^{\,(I)} \equiv a_{2} \ .
\label{isofluxes}
\eeqa
Isotropic fluxes are summarized in tables \ref{tableIsoNSRR} and \ref{tableIsoNon-Geometric}. 
\begin{table}[htb]
\begin{center}\begin{tabular}{|c|c|c|c||c|c|c|c|}
\hline
$\bar{F}_{---}$ & $\bar{F}_{|--}$ & $\bar{F}_{-||}$ & $\bar{F}_{|||}$ & 
$\bar{H}_{---}$ & $\bar{H}_{|--}$ & $\bar{H}_{-||}$ & $\bar{H}_{|||}$\\
\hline
\hline
$a_{3}$ & $a_{2}$ & $a_{1}$ & $a_{0}$ & $b_{3}$ & $b_{2}$ & $b_{1}$ & $b_{0}$ \\
\hline
\end{tabular}\end{center}
\caption{NS and RR isotropic fluxes. }
\label{tableIsoNSRR}
\end{table}

\begin{table}[htb]
\begin{center}\begin{tabular}{|c|c|c|c|c|c|c|c|}
\hline
$Q_{-}^{--}$ & $Q_{|}^{|-}$ & $Q_{|}^{-|}$ & $Q_{|}^{--}$ & $Q_{-}^{||}$ & $Q_{-}^{|-}$ & $Q_{-}^{-|}$ & $Q_{|}^{||}$\\
\hline
\hline
$\tilde{c}_{1}$ & $ \hat{c}_{1}$ & $ \check{c}_{1}$ & $ c_{0}$ & $c_{3}$ & 
$\check{c}_{2}$ & $\hat{c}_{2}$ & $\tilde{c}_{2}$\\
\hline
\end{tabular}\end{center}
\caption{Non-geometric isotropic fluxes.}
\label{tableIsoNon-Geometric}
\end{table}

The Ansatz of isotropic fluxes is compatible with vacua in which the geometric moduli are also isotropic, namely
\beq
U_{1}=U_{2}=U_{3}\equiv U  \quad ; \quad  T_{1}=T_{2}=T_{3}\equiv T \ .
\label{isoUT}
\eeq
This means, that there is only one overall complex structure modulus $U$ and one K\"ahler modulus $T$.
The model also includes the axiodilaton. In this case, the K\"ahler potential and the superpotential
reduce to 
\beqa
K & = & -3\,\log\left( -i\,(U-\bar{U})\right)  - \,\log\left( -i\,(S-\bar{S})\right)  - 
3\,\log\left( -i\,(T-\bar{T})\right) \nonumber \\[2mm]
W & = &P_{1}(U) + P_{2}(U)\,S + P_{3}(U)\,T \quad .
\label{kwiso}
\eeqa
The $P$'s are now cubic polynomials in the single complex structure moduli. They are given by 
\beqa
P_{1}(U) & = & a_{0}-3\,a_{1}\,U+3\,a_{2}\,U^{2}-a_{3}\,U^{3} \ ,
\label{P1Iso} \\[2mm]
P_{2}(U) & = & -b_{0}+3\,b_{1}\,U-3\,b_{2}\,U^{2}+b_{3}\,U^{3} \ ,
\label{P2Iso} \\[2mm]
P_{3}(U) & = & 3\, \left(  c_{0}+ (\hat{c}_{1}+\check{c}_{1}-\tilde{c}_{1}) \,U - 
(\hat{c}_{2}+\check{c}_{2}-\tilde{c}_{2})\,U^{2} - c_{3}\,U^{3} \right) \ .
\label{P3Iso}
\eeqa
This is the model considered in \cite{stw1,stw2}.

\subsection{Bianchi identities and tadpoles}
\label{ssec:bianchi}

The NSNS and generalized fluxes that follow from the T-duality chain can be regarded as structure
constants of an extended symmetry algebra of the compactification \cite{stw1, dabholkar}. This algebra includes isometry
generators $Z_{a}$ as well as gauge symmetry generators $X^{a}$, $a=1,\ldots, 6$, 
coming from the reduction of the $B$-field on $T^{6}$ with fluxes. We are interested in type IIB with O3/O7-planes where geometric and $R$-fluxes
are forbidden. In this case the algebra is given by 
\beqa
\left[ X^{a} , X^{b} \right]&=&Q_{c}^{ab}\,X^{c} \ , \nonumber \\
\left[ Z_{a} , X^{b} \right]&=&Q_{a}^{bc}\,Z_{c} \ , \label{zxalgebra}\\
\left[ Z_{a} , Z_{b} \right]&=&\bar H_{abc}\,X^{c}  \ . \nonumber
\eeqa
Notice that the $X^a$ span a 6-dimensional subalgebra in which the non-geometric $Q_c^{ab}$ are the structure constants.

Computing the Jacobi identities of the full 12-dimensional algebra we obtain the constraints 
\beq
\label{BianchiGen}
\bar H_{x[bc}\,Q^{ax}_{d]} =0  \qquad ; \qquad Q_{x}^{[ab}\,Q^{c]x}_{d}=0 \ .
\eeq
In the following we will refer to these identities in the shorthand notation $\bar H Q=0$ and
$Q Q=0$. The constraints on the fluxes can also be interpreted in terms of a nilpotency condition
$\cd^2=0$ on the operator $\cd=H\wedge + Q\cdot$ introduced in \cite{stw2}.

The RR fluxes are also constrained by Bianchi identities of the type $\cd \bar F=\cs$, where $\cs$
is a generalized form due to sources that are assumed smeared instead of localized. These Bianchi identities can be
understood as tadpole cancellation conditions on the RR 4-form $C_4$ and $C_8$ that couple to the sources.
The sources are just the orientifold O3/O7-planes and D3/D7-branes that can be present. In the IIB
orientifold that we are considering there is a flux-induced $C_4$ tadpole due to the coupling 
\beq
\int_{\M_{4} \times \M_{6}} C_{4} \wedge \bar H_{3} \wedge \bar F_{3} \ .
\label{c4tad}
\eeq 
There are further $C_4$ tadpoles due to O3-planes and to D3-branes that can also be added.
The total orientifold charge is -32, equally distributed among 64 O3-planes located at the fixed points of the 
orientifold involution $\sigma$. Each D3-brane has charge $+1$ and if they are located in the bulk, 
as opposed to fixed points of $\Z_2^3$, images must be included. Adding the sources to the flux
tadpole (\ref{c4tad}) leads to the cancellation condition
\beq
\label{O3tad}
a_{0}\,b_{3} - a_{1}^{(K)}\,b_{2}^{(K)} + a_{2}^{(K)}\,b_{1}^{(K)} - a_{3}\,b_{0}=N_{3} \ ,
\eeq 
where $N_3=32-N_{\rm D3}$ and $N_{\rm D3}$ is the total number of D3-branes. 

The non-geometric and RR fluxes can also combine to produce a tadpole for the RR $C_8$ form. The
contraction $Q \bar F_3$ is a 2-form and the flux-induced tadpole is due to the coupling
\beq
\int_{\M_{4} \times \M_{6}} C_{8} \wedge (Q \bar F_{3})
\label{c8tad}
\eeq 
Expanding the 2-form $(Q \bar{F}_{3})$ in the basis of 2-forms $\omega_I$, $I=1,2,3$,  yields coefficients
\beq
( Q \bar{F}_{3})_{I}=a_{0}\,c_{3}^{(I)}+a_{1}^{(K)}\,\cc_{2}^{(K I)} - a_{2}^{(K)}\,\cc_{1}^{(K I)}-a_{3}\,c_{0}^{(I)}  
\quad ; \quad  I=1,2,3 \ .
\label{o3d3tad}
\eeq
This means that there are induced tadpoles for $C_8$ components of type
$C_8 \sim d{\rm vol}_4 \wedge \widetilde{\omega}^I$, where $d{\rm vol}_4$ is the space-time volume 4-form
and $\widetilde{\omega}^I$ is the 4-form dual to $\omega_I$.
On the other hand, there are also $C_8$ tadpoles due to \mbox{O$7_I$-planes} that
have a total charge $+32$ for each $I$.  As discussed before, due to the orbifold
group $\Z_2 \times \Z_2$, there are \mbox{O$7_I$-planes} located at the 4 fixed tori of
$\sigma \theta_I$, where $\theta_I$ are the three order-two elements of $\Z_2 \times \Z_2$.
In the end we find the three tadpole cancellation conditions
\beq
\label{O7tad}
a_{0}\,c_{3}^{(I)}+a_{1}^{(K)}\,\cc_{2}^{(K I)} - a_{2}^{(K)}\,\cc_{1}^{(K I)}-a_{3}\,c_{0}^{(I)}
=N_{7_I}  \quad ; \quad  I=1,2,3 \ , 
\eeq
where $N_{7_I}=-32+N_{{\rm D}_{7_I}}$ and $N_{{\rm D}_{7_I}}$ is the number of 
\mbox{D$7_I$-branes} that are generically allowed.

In this work we mostly consider isotropic fluxes so that we will again make the Ansatz (\ref{isofluxes}).
Jacobi identities as well as tadpoles cancellation conditions become simpler. Computing $QQ=0$ constraints 
from (\ref{BianchiGen}) leave us with
\beqa
\label{BianchiXhatcheck}
\hat{c}_{2}\,\tilde{c}_{1} - \tilde{c}_{1}\,\check{c}_{2} + \check{c}_{1}\,\hat{c}_{2} - c_{0}\,c_{3}=0  \quad & ; & \quad 
c_{3}\,\tilde{c}_{1} - \check{c}_{2}^{2} + \tilde{c}_{2}\,\hat{c}_{2} - \hat{c}_{1}\,c_{3}=0 \ , \nonumber \\
c_{3}\,c_{0} - \check{c}_{2}\,\hat{c}_{1} + \tilde{c}_{2}\,\check{c}_{1} - \hat{c}_{1}\,\tilde{c}_{2}=0   \quad & ; & \quad
c_{0}\,\tilde{c}_{2} - \check{c}_{1}^{2} + \tilde{c}_{1}\,\hat{c}_{1} - \hat{c}_{2}\,c_{0}=0  \ ,
\eeqa
plus one additional copy of each condition with $\check{c}_{i} \leftrightarrow \hat{c}_{i}$.  
An important result is that saturating\footnote{This can be done using a computational algebra program as Singular \cite{singular} 
and solving over the real field. In \cite{stw1}, an analogous result is obtained manipulating this set of polynomial
constraints by hand.}  this ideal  with respect to the conditions $\check{c}_{i}  \not=  \hat{c}_{i}$ 
automatically implies that $\tilde{c}_{i}$ is complex. Therefore, it must be that
\beq
\check{c}_{1}=\hat{c}_{1} \equiv c_{1} \qquad ; \qquad 
\check{c}_{2}=\hat{c}_{2} \equiv c_{2}  \ .
\label{oneci}
\eeq
The cubic polynomial that couples the complex structure and K\"ahler moduli, c.f. (\ref{P3Iso}),  then reduces to
\beq
\label{P3Iso2}
P_{3}(U)=3\, \left( c_{0}+ (2\,c_{1}-\tilde{c}_{1}) \,U - (2\,c_{2}-\tilde{c}_{2})\, U^{2} - c_{3}\,U^{3} \right) \ .
\eeq   
Recall that the non-geometric fluxes are integer parameters. 

Upon using (\ref{oneci}), the Jacobi constraints satisfied by the non-geometric fluxes become
\beqa
\label{BianchiC}
c_0 \left(c_2-\tilde{c}_2\right)+ c_1\,(c_1-\tilde{c}_1) &=&0 \ , \nonumber \\
c_2\,(c_2-\tilde{c}_2)+c_3 \left(c_1-\tilde{c}_1\right)  &=&0  \ , \\
c_0 c_3-c_1 c_2 &=&0  \ . \nonumber 
\eeqa
This system of equations is easy to solve explicitly. The solution variety has three disconnected
pieces of different dimensions. The first piece has dimension four and it is characterized by fluxes
\beq
\begin{array}{lclcl}
c_{3}= \lambda_{p}\,k_2 & \quad ; \quad & c_{2}= \lambda_{p}\,k_1 & \quad ; \quad &  
\tilde{c}_{1}= \lambda_{q}\,k_2 + \lambda k_{1}  \quad ; \\
c_{1}= \lambda_{q}\,k_2 & \quad ; \quad & c_{0}= \lambda_{q}\,k_1 & \quad ; \quad & 
\tilde{c}_{2}= \lambda_{p}\,k_1 - \lambda k_{2} \quad .
\end{array}
\label{PieceA}
\eeq
Here $\lambda=1$, $(k_1,k_2)$ are two integers not zero simultaneously,  
and $(\lambda_p, \lambda_q)$ are two rays given by
\beq
\lambda_{p}=1+\frac{p}{\msm{\rm GCD}(k_{1},k_{2})} \quad ; \quad
\lambda_{q}=1+\frac{q}{\msm{\rm GCD}(k_{1},k_{2})}  \ ,
\label{raylambda}
\eeq
where $p, \, q, \in \Z$. By convention ${\rm GCD}(n,0)=|n|$. 
With coefficients given by the fluxes (\ref{PieceA}) the polynomial $P_3(U)$ turns out to factorize as
\beq
P_{3}(U)=3\, (k_1 + k_2 \,U) \,(\lambda_q -\lambda \, U - \lambda_{p}\,U^{2}) \ . 
\label{p3fact}
\eeq
Notice that we have taken into account that the non-geometric fluxes are integers.
The second piece of solutions is three dimensional, the set of fluxes can still be characterized
by (\ref{PieceA}) and $P_3(U)$ by (\ref{p3fact}), but with $\lambda\equiv 0$ and $\lambda_p\equiv1 $. 
Finally, the third piece has only two dimensions with fluxes and $P_3(U)$ specified by setting 
$\lambda\equiv 0$, $\lambda_p\equiv 0$ and $\lambda_q\equiv 1$.

As a byproduct of the above analysis we have isolated the real root of $P_3(U)$ that always exist.
In the next section we will explain how the nature of the remaining two roots is correlated with
the type of algebra fulfilled by the $X^a$ generators. For example, we will see that in the third piece
of solutions with $k_2=0$, the algebra is nilpotent.  

Let us now consider the constraints $\bar H Q=0$ that mix non-geometric and NSNS fluxes. 
Inserting the isotropic fluxes in (\ref{BianchiGen}), and using (\ref{oneci}),  we find  
\beqa
\label{BianchiB}
b_2 c_0-b_0 c_2+b_1(c_1- \tilde{c}_1) &=&0 \ , \nonumber \\
b_3 c_0-b_1 c_2+b_2 (c_1-\tilde{c}_1)&=&0 \ , \nonumber \\
b_2 c_1-b_0 c_3-b_1(c_2-\tilde{c}_2) &=&0   \ , \\
b_3 c_1-b_1 c_3-b_2 (c_2-\tilde{c}_2)&=&0 \ . \nonumber 
\eeqa
These conditions restrict the NSNS fluxes $b_A$ that determine the coupling between the
complex structure and the dilaton moduli through the polynomial $P_2(U)$ in (\ref{P2Iso}). 
In the next section we will discuss solutions to the full set of constraints that will lead
to specific forms for the polynomials $P_2(U)$ and $P_3(U)$.

The tadpole cancellation relations also become simpler in the isotropic case. In particular, 
the three constraints in (\ref{O7tad}), depending on $I$, reduce to just one condition.
Substituting the isotropic Ansatz and (\ref{oneci}) we obtain
\beq
 \label{O3tadIso}
 a_{0}\,b_{3}-3\,a_{1}\,b_{2}+3\,a_{2}\,b_{1}-a_{3}\,b_{0}=N_{3} \ ,
 \eeq
 \beq
 \label{O7tadIso}
 a_{0}\,c_{3}+a_{1}\,(2\,c_{2} -\tilde{c}_{2})-a_{2}\,(2\,c_{1} -\tilde{c}_{1})-a_{3}\,c_{0}=N_{7} \ .
 \eeq
These conditions constraint the RR fluxes. We consider the net O3/D3 and O7/D7 charges, $N_3$ and $N_7$,
to be free parameters.

\section{Algebras and fluxes}
\label{sec:alg}

In this section we discuss solutions to the Jacobi identities satisfied by the NSNS and
the non-geometric $Q$ fluxes. The key idea is twofold. First, the generators $X^a$ in (\ref{zxalgebra}) span 
a six-dimensional subalgebra whose structure constants are precisely the $Q^{ab}_c$. Second, when these 
fluxes are invariant under the $\Z_2^3$ symmetry described in section \ref{ssec:fluxes},
this subalgebra is rather constrained. We expect only a few subalgebras to be allowed and our
strategy is to identify them. In this way we will manage to provide explicit parametrizations
for non-geometric fluxes that satisfy the identity $QQ=0$. Once this is achieved, we will also be able to 
find the corresponding NSNS fluxes that fulfill $\bar H Q=0$.

We want to consider in detail the set of isotropic non-geometric fluxes given in
table \ref{tableIsoNon-Geometric} plus the conditions $\check{c}_{1}=\hat{c}_{1} \equiv c_{1}$,  
$\check{c}_{2}=\hat{c}_{2} \equiv c_{2}$. In this case the subalgebra simplifies to 
\beqa
\left[X^{2I-1}, X^{2J-1}\right] & = & \epsilon_{IJK} \left( \tilde c_1 \, X^{2K-1} + c_0\, X^{2K}\right)
\nonumber \ , \\
\left[ X^{2I-1}, X^{2J}\right] & = & \epsilon_{IJK} \left(c_2 \, X^{2K-1} + c_1\, X^{2K}\right)
\label{subiso} \ , \\
\left[ X^{2I}, X^{2J}\right] & = & \epsilon_{IJK} \left(c_3 \, X^{2K-1} + \tilde c_2 \, X^{2K}\right)
\nonumber \ ,
\eeqa
where $I,J,K=1,2,3$.
The Jacobi identities of this algebra are given in (\ref{BianchiC}).
To reveal further properties, it is instructive to compute the Cartan-Killing metric, denoted
$\cam$, with components 
\beq
\cam^{ab}= Q_{c}^{ad} \,\,Q_{d}^{bc} \ .
\label{ckmetric}
\eeq
For the above algebra of isotropic fluxes we find that the six-dimensional matrix $\cam$ is block-diagonal, 
namely
\beq
\cam = \diag(\cx_2, \cx_2, \cx_2) \ .
\label{mkcblock}
\eeq
The $2\times 2$ matrix $\cx_2$ turns out to be
\beq
\cx_2=-2\,\left(
\begin{array}{ll}
\qquad \tilde c_1^2 + 2 c_0c_2 + c_1^2  &  \tilde c_1 c_2 + c_1 c_2 + c_0 c_3 +  c_1 \tilde c_2     \\
\tilde c_1 c_2 + c_1 c_2 + c_0 c_3 + c_1 \tilde c_2  & \qquad \tilde c_2^2 + 2 c_1c_3 + c_2^2  \\
\end{array}
\right)
\ .
\label{mkc2}
\eeq
Since $\cx_2$ is symmetric, we conclude that $\cam$ can have up to two distinct
real eigenvalues, each with multiplicity three.

The full 12-dimensional algebra also enjoys distinctive features.
In the isotropic case the remaining algebra commutators involving NSNS fluxes are given by
\beqa
\left[Z_{2I-1}, Z_{2J-1}\right] & = & \epsilon_{IJK} \left(b_3 \, X^{2K-1} + b_2\, X^{2K}\right)
\ , \nonumber \\
\left[ Z_{2I-1}, Z_{2J}\right] & = & \epsilon_{IJK} \left(b_2 \, X^{2K-1} + b_1\, X^{2K}\right)
\ , \label{zzxiso} \\
\left[ Z_{2I}, Z_{2J}\right] & = & \epsilon_{IJK} \left(b_1 \, X^{2K-1} + b_0 \, X^{2K}\right)
\ . \nonumber 
\eeqa
The mixed piece of the algebra is determined by the non-geometric fluxes as
\beqa
\left[Z_{2I-1}, X^{2J-1}\right]  & = &  \epsilon_{IJK} \left(\tilde c_1 \, Z_{2K-1} + c_2\, Z_{2K}\right)
\ , \nonumber \\
\left[Z_{2I-1}, X^{2J}\right]  & = &  \epsilon_{IJK} \left(c_2 \, Z_{2K-1} + c_3\, Z_{2K}\right)
\ , \nonumber \\
\left[ Z_{2I}, X^{2J-1}\right]  & = &  \epsilon_{IJK} \left(c_0 \, Z_{2K-1} + c_1\, Z_{2K}\right)
\ , \label{zxziso} \\
\left[ Z_{2I}, X^{2J}\right]  & = &  \epsilon_{IJK} \left(c_1 \, Z_{2K-1} + \tilde c_2\, Z_{2K}\right)
\ . \nonumber 
\eeqa
Besides the Jacobi identities purely involving non-geometric fluxes, there are the additional mixed constraints
(\ref{BianchiB}).

Computing the full Cartan-Killing metric, denoted $\cam_{12}$, shows that there are no mixed $XZ$ terms. In fact, 
the matrix is again block-diagonal
\beq
\cam_{12} = \diag(\cx_2, \cx_2, \cx_2, \cz_2, \cz_2, \cz_2) \ ,
\label{fullmkcblock}
\eeq
with $\cx_2$ shown above. The new $2\times 2$ matrix $\cz_2$ is found to be 
\beq
\cz_2=-4\,\left(
\begin{array}{ll}
\quad b_3\tilde c_1  + 2 b_2c_2 + b_1 c_3  &   b_2(c_1+\tilde c_1) +b_1(c_2 + \tilde c_2)    \\
b_2(c_1+\tilde c_1) +b_1(c_2 + \tilde c_2)  &  \quad b_0\tilde c_2  + 2 b_1c_1 + b_2 c_0  \\
\end{array}
\right)
\ .
\label{fullmkc2}
\eeq
Here we have simplified using the Jacobi identities (\ref{BianchiB}).
We conclude that the allowed 12-dimensional algebras are such that the Cartan-Killing
matrix can have up to four distinct eigenvalues, each with multiplicity three.

Let us now return to the subalgebra spanned by the $X$ generators and the task of
solving the constraints (\ref{BianchiC}) that arise from the Jacobi identities $QQ=0$. 
The idea is to fulfill these constraints by choosing the non-geometric fluxes to be the
structure constants of six-dimensional Lie algebras whose Cartan-Killing matrix has the
simple block-diagonal form (\ref{mkcblock}). To proceed it is convenient to distinguish whether
$\cam$ is non-degenerate or not, i.e. whether the algebra is semisimple or not.
If \mbox{$\det \cam \! \not=\! 0$}, and $\cam$ is negative definite, the only possible algebra 
is the compact $\mathfrak{so(4)} \sim \mathfrak{su(2)^2}$. On the other hand, the only
non-compact semisimple algebra with the required block structure is $\mathfrak{so(3,1)}$. 
When $\det \cam\! =\! 0$, the algebra is non-semisimple. In this class to begin we find
two compatible algebras, namely the direct sum
$\mathfrak{su(2) + u(1)^3}$ and the semi-direct sum $\mathfrak{su(2) \oplus u(1)^3}$
that is isomorphic to the Euclidean algebra $\mathfrak{iso(3)}$.
The remaining possibility is that the non-semisimple algebra be completely solvable. One example is 
the nilpotent $\mathfrak{u(1)^6}$ that we disregard because the non-geometric fluxes vanish identically.
A second non-trivial solvable algebra, that is actually nilpotent,  will be discussed shortly.

After classifying the allowed 6-dimensional subalgebras the next step is to find the set of corresponding non-geometric fluxes.
Except for the nilpotent example, all other cases have an $\mathfrak{su(2)}$ factor.
This suggests to make a change of basis from $(X^{2I-1}, X^{2I})$, $I=1,2,3$,  to new generators
$(E^I, \widetilde E^I)$ such that basically one type, say $E^I$, spans $\mathfrak{su(2)}$. The 
$\Z_2^3$ symmetries of the fluxes require that we form combinations that transform in a 
definite way, For instance, $E^I$ can only be a combination of $X^{2I-1}$ and $X^{2I}$ with the
same $I$. Furthermore, for isotropic fluxes it is natural to make the same transformation for each $I$.
We will then make the $SL(2, \RR)$ transformation
\beq
\left(
       \begin{array}{c}
            E^I \\
            \widetilde{E}^I 
       \end{array}
\right)
= \frac{1}{|\Gamma |^{2}}
\left(
       \begin{array}{cc}
            -\alpha  &  \beta \\
            -\gamma  &\delta   
       \end{array}
\right)
\left(
       \begin{array}{c}
            X^{2I-1} \\
            X^{2I}    
       \end{array}
\right) \ ,
\label{chbasis}
\eeq
for all $I=1,2,3$. Here $|\Gamma|=\alpha\delta - \beta\gamma$, and it must be that $|\Gamma|\not=0$.
In the following we will refer to $(\a, \b, \g, \d)$ as the $\Gamma$ parameters.

Substituting in (\ref{subiso}) it is straightforward to obtain the algebra satisfied by the new generators
$E^I$ and $\wt E^J$. This algebra will depend on the non-geometric fluxes as well as on the parameters
$(\a, \b, \g, \d)$.
We can then prescribe the commutators to have the standard form for the allowed algebras
found previously. For instance, in the direct product examples we impose $\big[E^I, \wt E^J\big]=0$.

In the following sections we will discuss each compatible 6-dimensional algebra in more detail. The goal 
is to parametrize the non-geometric fluxes in terms of $(\a, \b, \g, \d)$. By construction these fluxes
will satisfy the Jacobi identities of the algebra. We will 
then solve the mixed constraints involving the NSNS fluxes. 
The main result will be an explicit factorization of the cubic polynomials $P_3(U)$
and $P_2(U)$ that dictate the couplings among the moduli.
 
\subsection{Semisimple algebras}

The algebra is semisimple when the Cartan-Killing metric is non-degenerate. This means
$\det \cam \not=0$ and hence $\det \cx_2 \not= 0$. Now, six-dimensional semisimple algebras
are completely classified. If $\cam$ is negative definite the algebra is compact so that
it must be $\mathfrak{so(4) \sim su(2) + su(2)}$. When $\cam$ has positive eigenvalues
the algebra is non-compact and it could be $\mathfrak{so(3,1)}$ or $\mathfrak{so(2,2)}$ but
the latter does not fit the required block-diagonal form (\ref{mkcblock}).

\subsubsection{$\mathfrak{so(4) \sim su(2)^2}$}
\label{subsubso4}

The standard commutators of this algebra are
\beq
\big[E^I, E^J\big]=\epsilon_{IJK} E^K \quad ; \quad
\big[\wt E^I, \wt E^J\big]=\epsilon_{IJK}\wt E^K \quad ; \quad \big[E^I, \wt E^J\big]=0 \ .
\label{su2su2}
\eeq
After performing the change of basis in (\ref{subiso}) we find that the non-geometric fluxes needed to describe this 
algebra can be parametrized as
\beq
\label{LimC}
\begin{array}{lcl}
c_{0}= \beta\, \delta \, (\beta+\delta) & \quad ; \quad & c_{3}=- \,\alpha\, \gamma \, (\alpha+\gamma) \quad , \\
c_{1} = \beta\, \delta \, (\alpha+\gamma) & \quad ; \quad & c_{2}=- \,\alpha\, \gamma \, (\beta+\delta) \quad , \\
\tilde{c}_{2}= \gamma^{2}\, \beta + \alpha^{2}\,\delta & \quad ; \quad & 
\tilde{c}_{1}=- \,(\gamma\, \beta^{2} + \alpha\,\delta^{2}) \quad , 
\end{array}
\eeq
provided that $|\Gamma|=(\a\d-\b\g) \not= 0$. It is easy to show that these fluxes 
verify the Jacobi identities (\ref{BianchiC}).
What we have done is to trade the six non-geometric fluxes, constrained by two independent conditions,
by the four independent parameters $(\alpha,\beta,\g, \d)$. These parameters are real but the resulting
non-geometric fluxes in (\ref{LimC}) must be integers.  

For future purposes we need to determine the cubic polynomial $P_3(U)$ that corresponds to the parametrized
non-geometric fluxes. Substituting in (\ref{P3Iso2}) yields
\beq
P_3(U)=3(\alpha U + \b)(\g U + \d)\big[(\a+\g)U + (\b+\d)\big] \ .
\label{p3so4}
\eeq
This clearly shows that in this case $P_3$ has three real roots. Moreover, the roots are all
different because $|\Gamma|\not=0$. We will prove that for other algebras $P_3$ has either
complex roots or degenerate real roots. The remarkable conclusion is that $P_3$ has three 
different real roots if and only if the algebra of the
non-geometric fluxes is the compact $\mathfrak{so(4) \sim su(2) + su(2)}$.
Alternatively, we may start with the condition that the polynomial has three different real roots
that we can choose to be at $0$, $-1$ and $\infty$ without loss of generality. These roots can then
be moved to arbitrary real locations by a linear fractional transformation
\beq
\cz = \frac{\a U + \b}{\g U + \d} \ .  
\label{zdef}
\eeq 
with $(\a, \b, \g, \d) \in \RR$ and $|\Gamma|\not=0$. By comparing the roots of $P_3$ in terms of
the fluxes with those in terms of the transformation parameters we rediscover the map (\ref{LimC})
and the associated $\mathfrak{su(2)^2}$ algebra. 
In the next sections we will see that the variable $\cz$ introduced above plays a very important physical r\^ole.

We now turn to the Jacobi constraints (\ref{BianchiB}) involving the NSNS fluxes.
Inserting the non-geometric fluxes (\ref{LimC}) we find that the $b_A$ can be completely fixed
by the $\Gamma$ parameters plus two new real variables $(\eps_1, \eps_2)$ as follows
\beqa
b_{0}&=&-\,(\epsilon_{1}\, \beta^{3} + \epsilon_{2}\, \delta^{3}) \ , \nonumber\\
b_{1}&=& \epsilon_{1}\, \alpha\,\beta^{2} + \epsilon_{2}\,\gamma\, \delta^{2} \ , \label{bso4} \\
b_{2}&=& -\,(\epsilon_{1}\, \alpha^{2}\,\beta + \epsilon_{2}\,\gamma^{2}\, \delta )\ , \nonumber \\
b_{3}&=&\epsilon_{1}\, \alpha^{3} + \epsilon_{2}\, \gamma^{3} \ . \nonumber
\eeqa
We also need to compute the polynomial $P_2(U)$ that depends on the NSNS fluxes. Substituting the above $b_A$
in (\ref{P2Iso}) yields
\beq
P_2(U)=\eps_1 (\alpha U + \b)^3 + \eps_2(\g U + \d)^3 \ .
\label{p2so4}
\eeq
It is easy to show that because $|\Gamma| \not= 0$, $P_2$ has complex roots whenever $\eps_1\eps_2\not=0$.
Contrariwise, $P_2$ has a triple real root if either $\eps_1$ or $\eps_2$ vanishes. 

We may expect that the full 12-dimensional algebra has special properties when $P_2$ has a triple root.
Indeed, inserting the fluxes in (\ref{fullmkc2}) yields $\det \cz_2 = 16 \eps_1\eps_2|\Gamma|^6$. 
Hence, the full Cartan-Killing matrix $\cam_{12}$ happens to be degenerate when $\eps_1\eps_2=0$.
To learn more about the full algebra it is convenient to switch from the original $Z_a$ generators to a new
basis $(D_I, \wt D_I)$ defined by
\beq
\left(
       \begin{array}{c}
            D_I \\
            \widetilde D_I 
       \end{array}
\right)
= \frac{1}{|\Gamma |^{2}}
\left(
       \begin{array}{cc}
            \d  &  \g \\
            \b  &\a   
       \end{array}
\right)
\left(
       \begin{array}{c}
            Z_{2I-1} \\
            Z_{2I}    
       \end{array}
\right) \ ,
\label{chzbasis}
\eeq
for $I=1,2,3$. It is straightforward to compute the piece of the full algebra 
generated by the $(D_I, \wt D_I)$. Substituting the parametrized fluxes in
(\ref{zzxiso}) and (\ref{zxziso}) we obtain 
\beq
\begin{array}{lcl}
\big[D_I, D_J\big]=-\eps_1 \, \epsilon_{IJK} E^K & \quad ; \quad & 
\big[\wt D_I, \wt D_J\big]= -\eps_2 \, \epsilon_{IJK}\wt E^K \quad , \\
\big[E^I, D_J\big]=\epsilon_{IJK} D_K & \quad ; \quad & 
\big[\wt E^I, \wt D_J\big]=\epsilon_{IJK} \wt D_K \quad . 
\end{array}
\label{morealg}
\eeq
All other commutators do vanish. 

A quick inspection of the whole algebra encoded in (\ref{su2su2}) and (\ref{morealg}) shows that when
either $\eps_1$, or $\eps_2$,  is zero, the $D_I$, or the $\wt D_I$, generate a 3-dimensional invariant Abelian 
subalgebra. Moreover, when  say $\eps_1=0$ and $\eps_2\not=0$, the  $\cz_2$ block of the 
full Cartan-Killing metric has one zero and one non-zero eigenvalue which is negative for $\eps_2 < 0$ 
and positive for $\eps_2 > 0$. The upshot is that when $\eps_1\eps_2=0$, the 12-dimensional algebra is  
$\mathfrak{iso(3) + g}$, where $\mathfrak{g}$ is either
$\mathfrak{so(4)}$ or $\mathfrak{so(3,1)}$. On the other hand, when $\eps_1 \eps_2 < 0$, the algebra is
$\mathfrak{so(4) + so(3,1)}$, whereas for $\eps_1, \eps_2 < 0$ it is $\mathfrak{so(4)^2}$, and
for $\eps_1, \eps_2 > 0$ it is $\mathfrak{so(3,1)^2}$. 

The methods developed in this section will be applied shortly to other subalgebras. In summary, the non-geometric and NSNS
fluxes can be parametrized using auxiliary variables $(\a,\b, \g, \d)$ and $(\eps_1, \eps_2)$ in such a way that 
the Jacobi identities are satisfied and flux-induced superpotential terms are explicitly factorized.
The full 12-dimensional algebras can be simply characterized after the changes of basis (\ref{chbasis}) and (\ref{chzbasis}) are performed.  

The auxiliary variables are constrained by the condition that the resulting fluxes be integers. 
This issue deserves further explanation. There are two cases depending on whether the
polynomial $P_2(U)$ has complex roots or not. If it does not, we can take $\epsilon_1=0$ to be concrete. From the
structure of the NSNS fluxes in (\ref{bso4}) it is then obvious that, for $\a \not=0$, the quotient $\b/\a$ is a rational number.
Going back to the non-geometric fluxes it can be shown that the ratios $\g/\a$ and $\d/\a$, as well as $\a^3$ and $\eps_2$ also
belong to $\mathbb{Q}$. If $P_2(U)$ admits complex roots the generic result is that $\eps_2/\eps_1$, $\b/\a$, $\a^3$, etc., involve
square roots of rationals. However, it happens that when at least one of the non-geometric parameters $(\a,\b, \g, \d)$ is zero
then all well defined quotients are again rational numbers.

\subsubsection{$\mathfrak{so(3,1)}$}

This is the Lorentz algebra. We can take $E^I$ to be the angular momentum, and $\wt E^J$ to be the
boost generators. Thus, the algebra can be written as
\beq
\big[E^I, E^J\big]=\epsilon_{IJK} E^K \quad ; \quad
\big[\wt E^I, \wt E^J\big]=-\epsilon_{IJK} E^K \quad ; \quad \big[E^I, \wt E^J\big]=\epsilon_{IJK} \wt E^K \ .
\label{s031}
\eeq
In this case the non-geometric fluxes that produce the algebra are found to be
\beq
\label{LimCSO31}
\begin{array}{lcl}
c_{0}=-\beta  \,\big(\beta ^2+\delta ^2\big) & \quad \ ; \quad &  
c_{3}=\,\alpha \, \big(\alpha ^2+\gamma ^2\big) \quad , \\
c_{1}= -\alpha  \,\big(\beta ^2+\delta ^2\big) & \quad ; \quad & 
c_{2}=\beta \, \big(\alpha ^2+\gamma ^2\big) \quad , \\
\tilde{c}_{2}= -\beta \, (\alpha ^2-\gamma ^2)-2\,\gamma \, \delta  \,\alpha  & \quad ; \quad &
\tilde{c}_{1}=\alpha  \big(\beta ^2-\delta ^2\big) + 2 \,\beta \, \gamma \, \delta \quad , 
\end{array}
\eeq
as long as  $|\Gamma| \not= 0$.  

Substituting the resulting non-geometric fluxes in (\ref{P3Iso2}) gives the $P_3(U)$ polynomial
\beq
P_3(U)=-3(\a U+\b)\big[(\alpha U + \b)^2 + (\g U + \d)^2 \big] \ .
\label{p3so31}
\eeq
Since $\Gamma\not=0$, $P_3$ always has complex roots. We will see that for non-semisimple algebras
all roots of $P_3$ are real, as for the compact $\mathfrak{so(4)}$. Hence, the important observation now is 
that $P_3$ has complex roots if and only if the algebra of the non-geometric fluxes is the non-compact 
$\mathfrak{so(3,1)}$.

The Jacobi constraints (\ref{BianchiB}) for the NSNS fluxes can again be solved in terms of the $\Gamma$ 
parameters plus two real constants that we again denote by $(\eps_1, \eps_2)$.  Concretely,
\beqa
b_{0}&=&-\beta\left(\beta^2 - 3\delta ^2\right) \epsilon _1 
-\delta \left(\delta ^2-3 \beta ^2\right) \epsilon _2 \ , \nonumber\\
b_{1}&=&  (\alpha \beta^2  - 2 \beta  \gamma  \delta - \alpha \delta^2) \epsilon _1
+\left(\gamma  \delta ^2 - 2 \alpha  \delta \beta - \gamma \beta^2\right) \epsilon _2 \ , 
\label{bso31}\\
b_{2}&=& \left(\beta  \gamma^2 + 2 \gamma  \delta  \alpha - \beta  \alpha^2\right) \epsilon _1
+\left(\delta  \alpha ^2+2 \beta \gamma  \alpha - \delta \gamma^2 \right) \epsilon _2 \ , \nonumber \\
b_{3}&=& \alpha\left(\alpha ^2 - 3\gamma ^2\right) \epsilon _1
+\gamma  \left(\gamma ^2-3 \alpha ^2\right) \epsilon _2  \ . \nonumber
\eeqa
These fluxes give rise to
\beq
P_2(U)=(\g U+\d)^3(\eps_1 \cz^3 - 3\eps_2 \cz^2 - 3 \eps_1 \cz + \eps_2) \ ,
\label{p2so31}
\eeq
where $\cz=(\a U + \b)/(\g U + \d)$ as before. The discriminant of this cubic polynomial is always
negative. Therefore, $P_2$ has three different real roots.

\subsection{Non-semisimple algebras}

In this case the algebra is the semidirect sum of a semisimple algebra and a solvable invariant subalgebra.
Lack of simplicity is detected imposing $\det \cam=0$ which requires $\det \cx_2=0$,
where $\cx_2$ is shown in (\ref{mkc2}).
Combining with the Jacobi identities (\ref{BianchiC}) we deduce that up to isomorphisms there are only
two solutions in which the solvable invariant subalgebra has dimension less than six. 
In practice this means that $\cx_2$ has only one zero eigenvalue. As expected from the 
underlying symmetries, this invariant subalgebra can only have dimension three and be $\mathfrak{u(1)^3}$.
The semisimple piece can only be $\mathfrak{su(2)}$. The two solutions are the direct and semidirect
sum discussed below.

The remaining possibility consistent with the symmetries is for the solvable invariant subalgebra 
to have dimension six. The criterion for solvability is that the derived algebra $\mathfrak{[g,g]}$ be
orthogonal to the whole algebra $\mathfrak{g}$ with respect to the Cartan-Killing metric.
In our case this means $Q^{ab}_c \cam^{dc}=0$, $\forall a,b,d$. The non-geometric fluxes 
further satisfy the Jacobi identities  $Q_{x}^{[ab}\,Q^{c]x}_{d}=0$. On the other hand, the stronger
condition for nilpotency is $\cam^{dc}=0$. For our algebra of isotropic fluxes given in (\ref{subiso}),
we find that all solvable flux configurations are necessarily nilpotent. The proof can be carried out
using the algebraic package {\it Singular} to manipulate the various ideals.  
This result is consistent with the fact that in our model $\cam$ is block-diagonal so that
when $\det \cam=0$, it has three or six null eigenvalues and in the latter situation $\cam$
is identically zero.
One obvious nilpotent algebra is $\mathfrak{u(1)^6}$, but it is uninteresting because the associated fluxes
vanish identically. There is a second solution described in more detail below. 

The allowed non-semisimple subalgebras can all be obtained starting from $\mathfrak{su(2)^2}$ and performing
contractions consistent with the underlying symmetries of the isotropic fluxes.
For example, setting $E^{\prime\, I} = E^I$, $\wt E^{\prime\, I} = \lambda \wt E^I$ in (\ref{su2su2})
and then letting $\lambda \to 0$ obviously gives the direct sum $\mathfrak{su(2)+ u(1)^3}$.
More generically we can take $E^{\prime\, I} = \lambda^a(E^I+ \wt E^I)$, 
$\wt E^{\prime\, I} = \lambda^b(E^I- \wt E^I)$, with $a\ge 0$, $b\ge 0$. The limit $a=0$, $b >0$,
$\lambda \to 0$ yields the Euclidean algebra $\mathfrak{iso(3)}$. Letting instead $2b=a >0$ and
contracting gives the nilpotent algebra.

In the coming sections we present the explicit configurations of non-geometric fluxes associated
to the non-semisimple subalgebras. The parametrization of NSNS fluxes is also computed. 
Evaluating the full 12-dimensional algebras in each case is straightforward.

\subsubsection{$\mathfrak{su(2)+ u(1)^3}$}

Since the algebra is a direct sum and one factor is Abelian,  
the brackets take the simple form
\beq
\big[E^I, E^J\big]=\epsilon_{IJK} E^K \quad ; \quad
\big[\wt E^I, \wt E^J\big]=0 \quad ; \quad \big[E^I, \wt E^J\big]=0 \ .
\label{su2d}
\eeq
Requiring that upon the change of basis the algebra (\ref{subiso}) is of this type
returns the following non-geometric fluxes 
\beq
\label{LimCFac}
\begin{array}{lcl}
c_{0}= \beta\, \delta^2  & \quad ; \quad & c_{3}=-\alpha\, \gamma^2\quad , \\
c_{1} = \beta\, \delta \, \gamma & \quad ; \quad & c_{2}= -\alpha\, \gamma \, \delta \quad , \\
\tilde{c}_{2}= \gamma^{2}\, \beta  & \quad ; \quad & 
\tilde{c}_{1}= -\alpha\,\delta^{2} \quad , 
\end{array}
\eeq
assuming $|\Gamma| \not= 0$. These fluxes automatically satisfy the Jacobi identities (\ref{BianchiC}).
They also satisfy the additional condition $c_0 c_2 = c_1 \tilde c_1$ arising from $\det \cx_2=0$.

The non-geometric fluxes of the algebra $\mathfrak{su(2)+ u(1)^3}$ lead to the $P_3(U)$ polynomial
\beq
P_3(U)=3(\a U+\b)(\g U + \d)^2 \ .
\label{p3su2d}
\eeq
Evidently, $P_3$ has one single and one double real root. 

The Jacobi identities $\bar H Q=0$ again fix the NSNS fluxes as in the previous cases.
The solution in terms of the free parameters is given by
\beqa
b_{0}&=&-\,(\epsilon_{1}\, \beta^{3} + \epsilon_{2}\, \delta^{3}) \ , \nonumber\\
b_{1}&=& \epsilon_{1}\, \alpha\,\beta^{2} + \epsilon_{2}\,\gamma\, \delta^{2} \ , \label{bsu2d}\\
b_{2}&=& -\,(\epsilon_{1}\, \alpha^{2}\,\beta + \epsilon_{2}\,\gamma^{2}\, \delta )\ , \nonumber \\
b_{3}&=&\epsilon_{1}\, \alpha^{3} + \epsilon_{2}\, \gamma^{3} \ . \nonumber
\eeqa
For the associated polynomial $P_2(U)$ we then find
\beq
P_2(U)=\eps_1 (\alpha U + \b)^3 + \eps_2(\g U + \d)^3 \ .
\label{p2su2d}
\eeq
As in the compact case, this $P_2$ has complex roots whenever $\eps_1 \eps_2 \not= 0$.

\subsubsection{$\mathfrak{su(2)\oplus u(1)^3 \sim iso(3)}$}

According to Levi's theorem, in general this algebra can be characterized as
\beq
\big[E^I, E^J\big]=\epsilon_{IJK} \big(E^K + \wt E^K \big) \quad ; \quad
\big[\wt E^I, \wt E^J\big]=0 \quad ; \quad \big[E^I, \wt E^J\big]=\epsilon_{IJK} \wt E^K \ .
\label{su2u13sd}
\eeq
The typical form of the Euclidean algebra in three dimensions is recognized
after the isomorphism $(E^I-\wt E^I) \to \widehat E^I$. 
The non-geometric fluxes needed to reproduce the above commutators turn out to be
\beq
\label{LimCFacSemi}
\begin{array}{lcl}
c_{0}=-\delta ^2\,(\beta-\delta) & \quad \ ; \quad &  
c_{3}=\gamma ^2\,(\alpha-\gamma) \quad , \\
c_{1}= -\delta ^2\,(\alpha-\gamma) & \quad ; \quad & 
c_{2}=\gamma ^2\,(\beta-\delta) \quad , \\
\tilde{c}_{2}= \gamma ^2\,(\beta+\delta)- 2\,\gamma \, \delta  \,\alpha  & \quad ; \quad &
\tilde{c}_{1}=-\delta^2\, (\alpha+\gamma)  + 2\, \gamma \, \delta \, \beta \quad , 
\end{array}
\eeq
for $|\Gamma| \not= 0$. Besides the Jacobi identities these fluxes satisfy $4 c_0c_2=-(c_1-\tilde c_1)^2$,
by virtue of $\det \cx_2=0$.

For the flux configuration of this algebra the $P_3(U)$ polynomial becomes
\beq
P_3(U)=3(\g U + \d)^2\big[(\g-\a)U + (\d-\b)\big] \ .
\label{p3su2sd}
\eeq
As in the direct sum $\mathfrak{su(2) + u(1)^3}$, $P_3$ has one single and one double real root. 

The NSNS fluxes can be determined from the Jacobi identities (\ref{BianchiB}).
Introducing again parameters $(\eps_1,\eps_2)$ leads to
\beqa
b_{0}&=&-\d^2  \, \left(\beta \, \epsilon _1+\delta \, \epsilon _2\right)  \ , \nonumber\\
b_{1}&=&   \msm{\frac{1}{3}} \, \delta (\alpha\, \delta + 2\, \beta  \,\gamma)\epsilon _1
+ \gamma \, \delta^2  \,\epsilon _2  \ , \label{bsu2sd} \\
b_{2}&=& - \msm{\frac{1}{3}} \gamma (\beta \, \gamma +2 \,\alpha \, \delta ) \epsilon _1
- \gamma^2 \, \delta \, \epsilon _2  \ , \nonumber \\
b_{3}&=&   \gamma ^2\, \left(\alpha \, \epsilon _1+\gamma \, \epsilon _2\right) \ ,  \nonumber
\eeqa
The companion polynomial $P_2(U)$ of NSNS fluxes is fixed as
\beq
P_2(U)=(\g U + \d)^2\left[\eps_1(\a U + \b)  + \eps_2(\g U + \d) \right] \ .
\label{p2su2sd}
\eeq
Analogous to the non-compact case, this $P_2$ has only real roots, but one of them is degenerate.

\subsubsection{Nilpotent algebra}

To search for flux configurations that generate a nilpotent algebra we impose that the
Cartan-Killing metric vanishes. Now, in our model $\cam=0$ implies the much simpler conditions 
$\det \cx_2=0$ and $\Tr \cx_2=0$.
Up to isomorphisms, we find only one non-trivial solution. This is the expected result based on
the known classification of 6-dimensional nilpotent algebras\footnote{ 
A table and references to the original literature are given in \cite{gmpt}.}.

{}From the 34 isomorphism classes of nilpotent algebras, besides $\mathfrak{u(1)^6}$, only
one is compatible with isotropic fluxes invariant under $\Z_2 \times \Z_2$. The algebra
is 2-step nilpotent and its brackets can be written as 
\beq
\big[E^I, E^J\big]= \epsilon_{IJK} \wt E^K  \quad ; \quad
\big[\wt E^I, \wt E^J\big]=0  \quad ; \quad \big[E^I, \wt E^J\big]=0 \ .
\label{nilal}
\eeq
Up to isomorphisms this is the algebra labelled $n(3.5)$ in Table 4 of \cite{gmpt}.

The change of basis from the original $(X^{2I-1}, X^{2I})$ generators to the $(E^I, \wt E^I)$
is still given by (\ref{chbasis}). 
Starting from the $X$ commutators in (\ref{subiso}) we can then deduce
fluxes such that the nilpotent algebra (\ref{nilal}) is reproduced. In this way we obtain 
\beq
\label{LimCNilp}
\begin{array}{lcl}
c_{0}= \delta ^3 & \quad ; \quad & c_{3}=- \gamma^3 \quad , \\
c_{1} = \delta^2\, \gamma  & \quad ; \quad & c_{2}=- \delta \, \gamma^2 \quad , \\
\tilde{c}_{2}= \delta \, \gamma^{2} & \quad ; \quad & 
\tilde{c}_{1}=- \delta^{2}\, \gamma  \quad . 
\end{array}
\eeq
Notice that these fluxes only depend on two independent parameters. This occurs because besides the
Jacobi constraints there are two more conditions $\det \cx_2=0$ and $\Tr \cx_2=0$.  
The non-geometric fluxes of the nilpotent algebra generate the $P_3(U)$ polynomial
\beq
P_3(U)=3(\g U+\d)^3 \ .
\label{p3nil}
\eeq
Clearly, $P_3$ always has one triple real root. 

In analogy with all previous examples, the $\bar H Q=0$ Jacobi identities determine the NSNS fluxes in terms of two
additional parameters $(\eps_1, \eps_2)$. Inserting the non-geometric fluxes of the nilpotent algebra 
in (\ref{BianchiB}) readily yields
\beqa
b_{0}&=&-\delta ^2 \left(\delta \,\epsilon _2+\gamma  \,\epsilon _1\right) \ ,   \nonumber\\
b_{1}&=&  \gamma \, \delta ^2\, \epsilon _2
-\msm{\frac{1}{3}} \,\delta \left(\delta^2 -2\, \gamma ^2\right) \epsilon _1  \ , \label {bnil}\\
b_{2}&=&  - \gamma^2\, \delta \, \epsilon _2
+ \msm{\frac13} \gamma \left(2\, \delta^2 - \gamma^2 \right) \epsilon _1  \ , \nonumber \\
b_{3}&=&   \gamma^2 \left(\gamma \,  \epsilon _2-\delta  \,\epsilon _1\right) \ . \nonumber
\eeqa
Substituting in (\ref{P2Iso}) we easily obtain the corresponding polynomial
\beq
P_2(U)=(\g U + \d)^2\left[\eps_2(\g U + \d)  + \eps_1(\g - \d U) \right] \ .
\label{p2nil}
\eeq
As in $\mathfrak{su(2) \oplus u(1)^3}$, this $P_2$ has one single and one double real root.
Without loss of generality we can choose $\a=-\d$ and $\b=\g$ in order to write $P_2$
in terms of the variable $\cz=(\a U + \b)/(\g U + \d)$ as 
\beq
P_2(U)=(\g U+\d)^3(\eps_1 \cz  + \eps_2) \ .
\label{p2nils}
\eeq
The advantage of this choice of parameters will become evident when we perform a transformation
from $U$ to $\cz$ in the scalar potential.

\section{New variables and RR fluxes}
\label{sec:newvars}

In type IIB orientifolds, the superpotential depends on the complex structure parameter $U$
through the three cubic polynomials $P_1(U)$, $P_2(U)$ and $P_3(U)$ induced respectively by
RR, NSNS and non-geometric $Q$-fluxes. Our results in last section show that
the last two polynomials can be concisely written as 
\beq
P_2(U)=(\g U + \d)^3 \cp_2(\cz) \qquad ; \qquad P_3(U)=(\g U + \d)^3 \cp_3(\cz) \ ,
\label{cp23def}
\eeq 
where $\cz=(\a U+ \b)/(\g U + \d)$. The real parameters $(\a, \b, \g, \d)$, with
\mbox{$|\Gamma|=(\a\d-\b\g)\not=0$}, encode the non-geometric fluxes. For the NSNS fluxes two additional
real constants $(\eps_1, \eps_2)$ are needed. As summarized in table \ref{tablecps}, $\cp_2(\cz)$ and 
$\cp_3(\cz)$ take very specific forms according to the subalgebra of the $Q$-fluxes. 

\begin{table}[htb]
\small{
\renewcommand{\arraystretch}{1.15}
\begin{center}
\begin{tabular}{|c|c|c|c|}
\hline
$Q$-subalgebra & $\cp_3(\cz)/3$ & $\cp_2(\cz)$ & $\cp_1(\cz)$ \\
\hline
\hline
$\mathfrak{so(4)}$ & $\cz(\cz+1)$ & $\eps_1 \cz^3 + \eps_2$ & $\xi_3(\eps_1-\eps_2\cz^3) + 3\xi_7\cz(1-\cz)$ \\
\hline
 & & & $\xi_3(\eps_1 +3\eps_2\cz-3\eps_1\cz^2-\eps_2\cz^3)$ \\
\raisebox{2.5ex}[0cm][0cm]{$\mathfrak{so(3,1)}$} & 
\raisebox{2.5ex}[0cm][0cm]{$-\cz(\cz^2+1)$} & 
\raisebox{2.5ex}[0cm][0cm]{$\eps_1 \cz^3 -3\eps_2 \cz^2 - 3 \eps_1 \cz + \eps_2$}  &  
$ + 3\xi_7(\cz^2+1)$ \\
\hline
$\mathfrak{su(2)+u(1)^3}$ & $\cz$ & $\eps_1 \cz^3 + \eps_2$ & $\xi_3(\eps_1-\eps_2\cz^3) - 3\xi_7\cz^2$  \\
\hline
$\mathfrak{su(2)\oplus u(1)^3}$ & $1-\cz$ & $\eps_1 \cz + \eps_2$ & 
$3\lambda_1 \cz + 3\lambda_2 \cz^2 + \lambda_3 \cz^3$  \\
\hline
$\mathfrak{nil}$ & $1$ & $\eps_1\cz  + \eps_2 $ & $3\lambda_1 \cz + 3\lambda_2 \cz^2 + \lambda_3 \cz^3$  \\
\hline
\end{tabular}
\end{center}
\caption{$Q$-subalgebras and polynomials}
\label{tablecps}
}
\end{table}

A very nice property of the variable $\cz$ is its invariance under the $SL(2,\Z)_U$ modular
transformations
\beq
U^\prime = \frac{k \, U + \ell}{m \, U + n} \quad ; \quad k,\, \ell, \, m , \, n \, \in \Z 
\quad ; \quad kn-\ell m=1 \ .
\label{umodt}
\eeq
Since this is a symmetry of the compactification, the effective action must be invariant.
The K\"ahler potential, $K=-3\log[-i(U-\bar U)] + \cdots$, clearly transforms as
\beq
K^\prime = K + 3 \log| m U + n|^2 \ .
\label{kmodt}
\eeq 
Therefore, the physically important quantity $e^K |W|^2$ is invariant as long as the
superpotential satisfies
\beq
W^\prime = \frac{W}{(m U + n)^3} \ .
\label{wmodt}
\eeq
In order for $W$ to fulfill this property the fluxes must transform in definite patterns.
In fact, it follows that (\ref{wmodt}) holds separately for each of the flux induced polynomial
$P_i(U)$.

We claim that the fluxes transform under $SL(2,\Z)_U$ precisely in such a manner that $\cz^\prime=\cz$.
The proof begins by first finding how the $Q$-fluxes mix among themselves
from the condition $P_3^\prime=P_3/(m U + n)^3$. For example, under $U^\prime=-1/U$, the non-geometric
fluxes transform as
\beq
c_0^\prime=-c_3 \quad , \quad c_1^\prime=c_2 \quad , \quad c_2^\prime=-c_1 \quad , \quad c_3^\prime=c_0 
\quad , \quad \tilde c_1^{\, \prime}= \tilde c_2 \quad , \quad  \tilde c_2^{\, \prime}= -\tilde c_1 \ .
\label{cduals}
\eeq      
Next we read off the corresponding transformation of
the parameters $(\a, \b, \g, \d)$ that are better thought of as the elements of a matrix $\Gamma$.
The result is
\beq
\Gamma^\prime=
\left(
\begin{array}{ll}
\a^\prime & \b^\prime \\
\g^\prime & \d^\prime
\end{array} \right) 
=\left(
\begin{array}{ll}
\a & \b \\
\g & \d
\end{array} \right)  \!\!
\left(
\begin{array}{cc}
n & \!\!\!\! -\ell \\
\!\!\!   -m & k
\end{array} \right) 
\label{Gmodt}
\eeq
It easily follows that $\cz^\prime=\cz$. Notice that $|\Gamma^\prime|=|\Gamma|$.

For the NSNS fluxes we can study the transformation of $P_2$ with coefficients given by the $b_A$. 
Alternatively, we may start from $P_2$ written as function of $\cz$ as in (\ref{cp23def}).
The conclusion is that the transformation of the $b_A$ is also determined by $\Gamma^\prime$
together with $(\eps_1^\prime,\eps_2^\prime)=(\eps_1, \eps_2)$. This is valid for all
$Q$-subalgebras.

At this point it must be evident that we want to change variables from $U$ to $\cz$. It is
also convenient to trade the axiodilaton $S$ and the K\"ahler modulus $T$ by new fields
defined by
\beq
\cs = S + \xi_s \quad ; \quad \ct=T + \xi_t \ ,
\label{csctdef}
\eeq
where the shifts $\xi_s$ and $\xi_t$ are some real parameters. The motivation is that such
shifts in the axions $\re S$ and $\re T$
can be reabsorbed into RR fluxes as explained in the following.   

\subsection{Parametrization of RR fluxes}
\label{ss:rr}

The systematic procedure is to express the RR fluxes $a_A$ in such a way that
their contribution to the superpotential is of the form
\beq
P_1(U)=(\g U + \d)^3 \widehat \cp_1(\cz) \ , 
\label{cp1hatdef}
\eeq 
in complete analogy with (\ref{cp23def}). 
To arrive at this factorization we must relate the four RR fluxes $a_A$ to the parameters
$(\a, \b, \g, \d)$ that define $\cz=(\a U+ \b)/(\g U + \d)$, and to four additional independent variables. 
Obviously, $\widehat \cp_1(\cz)$ can be expanded in the monomials $(1,\cz,\cz^2,\cz^3)$. However,
a more convenient basis contains the already known polynomials $\cp_3$ and $\cp_2$ that are
generically linearly independent. We still need two independent polynomials and these are taken to be 
the duals $\widetilde \cp_3$ and $\widetilde \cp_2$. The dual $\widetilde \cp$ is such that 
$\cp \to \widetilde \cp/\cz^3$ when $\cz \to -1/\cz$.
The last two subalgebras in table \ref{tablecps} must be treated slightly different because
linear independence of $\cp_3$ and $\cp_2$ fails for particular properties of the NSNS flux parameter
$\eps_1$.

We concretely make the expansion
\beq
\widehat \cp_1(\cz) = \xi_s \cp_2(\cz) + \xi_t \cp_3(\cz) + \cp_1(\cz) \ .
\label{cp1hatexp}
\eeq
In the full superpotential the first two terms in $\widehat \cp_1$ will precisely offset the axionic shifts 
in the new variables $\cs$ and $\ct$. Let us now discuss the remaining piece $\cp_1(\cz)$ that also
depends on the  $Q$-subalgebra and is displayed in table \ref{tablecps}.
As explained before, for the first three subalgebras in the table we can further choose
\beq
 \cp_1(\cz) = \xi_7 \widetilde \cp_3(\cz) - \xi_3 \widetilde \cp_2(\cz) \ .
\label{cp1defa}
\eeq
A motivation for this choice is that the RR tadpoles turn out to depend on the RR fluxes
only through the coefficients $(\xi_3, \xi_7)$.

For the last two subalgebras in table \ref{tablecps}, $\cp_3$ and $\cp_2$ are not independent when 
$\eps_1$ takes a particular critical value.
For $\mathfrak{su(2)\oplus u(1)^3}$ this happens when $\eps_1=-\eps_2$, whereas for
the nilpotent algebra the critical value is $\eps_1=0$. 
To take into account these possibilities, compensating at the same time
for the axionic shifts, we still make the decomposition (\ref{cp1hatexp}) but with
\beq
\cp_1(\cz) = 3\lambda_1 \cz + 3 \lambda_2 \cz^2  + \lambda_3 \cz^3 \ .
\label{cp1defb}
\eeq 
Away from the critical values of $\eps_1$ we can take $\lambda_1=0$ because $\xi_s$ and
$\xi_t$ are independent parameters. At the critical value necessarily $\lambda_1 \not=0$
but in this case $\xi_s$ and $\xi_t$ enter in the RR fluxes in only one linearly 
independent combination.  
The RR tadpoles happen to depend just on the parameters $(\lambda_2,\lambda_3)$. 

The next step is to compare the expansion of $P_1(U)$ in $U$ with its factorized form, c.f. (\ref{cp1hatdef}) and 
(\ref{P1Iso}). In this way we can obtain an explicit parametrization of the RR fluxes $a_A$ in terms of the
variables that determine $\widehat \cp_1(\cz)$, namely $(\xi_s,\xi_t)$ together with $(\xi_3,\xi_7)$ or
$(\lambda_1, \lambda_2, \lambda_3)$, depending on the $Q$-subalgebra.
These results are collected in the appendix. We stress that the $\xi$'s and $\lambda$'s are real
parameters but the emerging RR fluxes must be integers.

A vacuum solution in which the moduli $(\cz, \cs, \ct)$ are fixed generically requires specific
values of the non-geometric, NSNS and RR fluxes. These fluxes also generate RR tadpoles that
must be balanced by adding orientifold planes or D-branes. To determine the type of sources
that must be included we need 
to evaluate the RR tadpole cancellation conditions using all parametrized fluxes. 
Substituting in (\ref{O3tadIso}) and (\ref{O7tadIso}) we arrive at the very compact expressions for
the number of sources $N_3$ and $N_7$ gathered in table \ref{tabletads}.
As advertised before, the RR fluxes only enter either through the parameters $(\xi_3, \xi_7)$ or
$(\lambda_2,\lambda_3)$. The non-geometric and NSNS fluxes only contribute through
$|\Gamma|^3$ and $(\eps_1,\eps_2)$. We will see that there is also a clear correlation of the
tadpoles with the vevs of the moduli.

\begin{table}[htb]
\small{
\renewcommand{\arraystretch}{1.15}
\begin{center}
\begin{tabular}{|c|c|c|}
\hline
$Q$-subalgebra & $N_3/|\Gamma|^3$ & $N_7/|\Gamma|^3$ \\
\hline
\hline
$\mathfrak{so(4)}$ & \ \ $(\eps_1^2 + \eps_2^2)\,\xi_3$ & \, $2 \,\xi_7$ \\
\hline
$\mathfrak{so(3,1)}$ & \!\! $4(\eps_1^2 + \eps_2^2)\,\xi_3$ & \, $4 \,\xi_7$ \\
\hline
$\mathfrak{su(2)+u(1)^3}$ & \ \ $(\eps_1^2 + \eps_2^2)\, \xi_3$ & \!\! $\,\xi_7$ \\
\hline
$\mathfrak{su(2)\oplus u(1)^3}$ & $\lambda_2\, \eps_1-\lambda_3\, \eps_2$ & $\lambda_2 + \lambda_3$ \\
\hline
$\mathfrak{nil}$ & $\lambda_2 \,\eps_1 -\lambda_3\,\eps_2$  & $\lambda_3$   \\
\hline
\end{tabular}
\end{center}
\caption{$Q$-subalgebras and RR tadpoles}
\label{tabletads}
}
\end{table}

Finally, let us remark that, just like $(\eps_1, \eps_2)$, the $\xi$ and $\lambda$ variables are all invariant 
under modular transformations of the complex structure $U$. Indeed, from the explicit parametrization
of the RR fluxes $a_A$ we deduce that their correct behavior under $SL(2,\Z)_U$, analogous to
(\ref{cduals}), precisely follows from the transformation of $(\a, \b, \g, \d)$ in (\ref{Gmodt}).
This is of course consistent with the fact that the number of sources $N_3$ and $N_7$ in the tadpoles
are physical quantities that must be modular invariant.

\subsection{Moduli potential in the new variables}

We have just seen how a systematic parametrization of the fluxes has guided us to new moduli fields
denoted $(\cz, \cs, \ct)$. As we may expect, the effective action in the transformed variables also takes a form 
more suitable for finding vacua.  The shifts in the axionic real parts of the axiodilaton and the K\"ahler field
do not affect the K\"ahler potential $K$ whereas in the superpotential $W$ they can be reabsorbed in RR fluxes.
On the other hand, the change from the complex structure $U$ to $\cz$ is the $SL(2,\RR)$ transformation
$U=(\b-\d\cz)/(\g\cz-\a)$ whose effect on $K$ and $W$ is completely analogous to a modular transformation
except for factors of $|\Gamma|=(\a\d-\b\g)$. Combining previous results we obtain $e^K|W|^2 \to e^\ck|\cw|^2$, 
where the transformed K\"ahler potential $\ck$ and superpotential $\cw$ are given by
\beqa
\mathcal{K} & = &-3 \,\log\left( -i\,(\mathcal{U}-\bar{\mathcal{U}})\right)  - 
\,\log\left( -i\,(\mathcal{S}-\bar{\mathcal{S}})\right)  - 3 \,\log\left(- i\,(\mathcal{Z}-\bar{\mathcal{Z}}) \right)
\ , \label{KModular} \\[2mm]
\mathcal{W}  & = &  |\Gamma|^{3/2} \left[\ct \, \cp_3(\cz) \,+ \, \cs \, \cp_2(\cz)
+ \cp_1(\cz) \right] \ . 
\label{WModular}
\eeqa             
The flux-induced polynomials $\cp_i(\cz)$ are displayed in table \ref{tablecps} for each $Q$-subalgebra.  
In the effective 4-dimensional action with \neq1 supergravity  
the functions $\ck$ and $\cw$ determine the scalar potential of the moduli according to
\beq
V = e^\ck \left\{ \sum_{\Phi=\cz,\cs,\ct} \ck^{\Phi\bar \Phi} |D_\Phi \cw|^2 - 3|\cw|^2 \right\} \ .
\label{VModular}
\eeq
We are interested in supersymmetric minima for which $D_\Phi \cw = \partial_\Phi \cw + \cw \partial_\Phi \ck =0$,
for all fields.

\section{Supersymmetric vacua}
\label{sec:vac}

This section is devoted to searching for supersymmetric vacua of the moduli potential induced
by RR, NSNS and non-geometric fluxes together. We will show that by using our new variables
the problem simplifies substantially and analytic solutions are feasible. 

Supersymmetric vacua are characterized by the vanishing of the F-terms. In our setup the 
conditions are
\beqa
D_{\ct}\mathcal{W}&=&\frac{\partial \cw}{\partial \ct} + 
\frac{3i \cw}{2 \im \ct}=0 \ , \nonumber \\[2mm]
D_{\cs}\mathcal{W}&=&\frac{\partial \cw}{\partial \cs} + 
\frac{i \cw}{2 \im \cs}=0  \ , \label{FFlat} \\[2mm]
D_{\cz}\mathcal{W}&=&\frac{\partial \cw}{\partial \cz} + 
\frac{3i \cw}{2 \im \cz}=0  \ . \nonumber 
\eeqa
The task is to determine whether there are solutions with moduli completely stabilized at vevs
denoted 
\beq
\cz_0 = x_0 + i y_0 \quad ; \quad \cs_0 = s_0 + i\sigma_0 \quad ; \quad \ct_0 = t_0 + i \mu_0 \ .
\label{vzst}
\eeq
The vacua are either Minkowski or AdS because the potential
(\ref{VModular}) at the minimum is given by $V_0 = - 3 e^{\ck_0} |\cw_0|^2 \leq 0$.

Besides stabilization, there are further physical requirements. At the minimum
the imaginary part of the axiodilaton, $\sigma_0$, must be positive for the reason it is the inverse 
of the string coupling constant $g_s$.
It can be argued that the geometric moduli are subject to similar conditions. The main assumption is
that they arise from the metric of the internal space, which is $\T^6$ in absence of fluxes.
In particular, the K\"ahler modulus has $\im \ct = e^{-\phi} A$, where $A$ is
the area of a 4-dimensional subtorus. Hence, it must be $\mu_0 > 0$.  
Notice also that the internal volume is measured by $V_{int}=(\mu_0/\sigma_0)^{3/2}$. 
For the transformed complex structure $\cz$ it happens that $\im \cz = |\Gamma| \im U/|\g U + \d|^2$.
Therefore, necessarily $\im \cz_0 = y_0 \not= 0$ because for $\im U_0=0$ the internal space is degenerate.
Without loss of generality we choose that $\im U_0$ is always positive.

Another physical issue is whether the moduli take values such that the effective supergravity action
is a reliable approximation to string theory. Specifically, the string coupling $g_s = 1/\sigma_0$ is expected to be small
to justify the exclusion of non-perturbative string effects. Conventionally, there is also a requirement
of large internal volume to disregard corrections in $\alpha^\prime$. 
However, in presence of non-geometric fluxes the internal space might be
a T-fold in which there can exist cycles with sizes related by T-duality \cite{hull, dabholkar}. Thus, for large
volume there could be tiny cycles whose associated winding modes would be light. To date these effects
are not well understood. At any rate, in this work we limit ourselves to finding supersymmetric vacua
of an effective field theory defined by a very precise K\"ahler potential and flux-induced superpotential. 
A more detailed discussion of the landscape of vacua is left for section \ref{sec:lands}. We will see that the
moduli can be fixed at small string coupling and cosmological constant.   

In the following we will first consider supersymmetric Minkowski vacua that have 
\mbox{$\cw=0$} at the minimum.
In our approach it is straightforward to show that for isotropic fluxes such vacua are disallowed.
We then turn our attention to the richer class of ${\rm AdS}_4$ vacua.
Since superpotential terms adopt very specific forms depending 
on the particular subalgebra satisfied by the non-geometric fluxes, we will study
the corresponding vacua case by case.
We will mostly focus on the model associated to the non-geometric fluxes of the compact
$\mathfrak{su(2)^2}$ but will also consider other allowed subalgebras to some extent.

\subsection{Minkowski vacua}

Minkowski solutions with zero cosmological constant require that the potential vanishes. Imposing supersymmetry
further implies that the superpotential must be zero at the minimum $(\cz_0, \cs_0, \ct_0)$.
A key property of the superpotential (\ref{WModular}) is its linearity in $\cs$ and $\ct$. This implies
in particular that the F-flat conditions $D_\cs\cw=0$ and $D_\ct\cw=0$, together with $\cw=0$, reduce just to
\beq
\cp_3(\cz_0)=\cp_2(\cz_0)=\cp_1(\cz_0)=0 \ .
\label{minkcon}
\eeq
The third condition $D_\cz\cw=0$ yields a linear relation between $\cs_0$ and $\ct_0$ so that not all moduli
can be stabilized. The situation is actually worse because (\ref{minkcon}) cannot be fulfilled appropriately. Indeed, 
for the specific polynomials for each subalgebra shown in table \ref{tablecps}, it is evident that $\cp_3$ and $\cp_2$ 
can only have a common real root $\cz_0$. But then $\im U_0=\im \cz_0=0$ and this is inconsistent with a well defined 
internal space. 

It must be emphasized that we are assuming that non-geometric fluxes, and their induced $\cp_3$, are non-trivial.
Our motivation is to fix the K\"ahler modulus without invoking non-perturbative effects.
If only RR and NSNS fluxes are turned on there do exist physical supersymmetric Minkowski vacua in which only the
axiodilaton and the complex structure are stabilized \cite{kst, dgkt}. In such solutions the RR and NSNS fluxes
must still satisfy a non-linear constraint \cite{dgkt, gray}.

No-go results for supersymmetric Minkowski vacua in presence of  non-geometric fluxes have been obtained previously
\cite{acfi, tasinato, gray}
\footnote{In \cite{tasinato} it is further shown that Minkowski vacua with all moduli
stabilized can exist in more general setups having more complex structure than K\"ahler moduli (in IIB language).}. 
In \cite{acfi} the existence was disproved supposing special solutions for the Jacobi identities (\ref{BianchiC}). 
We are now extending the proof to all possible non-trivial {\it isotropic} non-geometric fluxes solving these constraints.

\subsection{${\rm AdS}_4$ vacua}
\label{sub:ads}

We now want to solve the supersymmetry conditions when $\cw \not=0$. The three complex
equations $D_\Phi\cw=0$, $\Phi=\cz,\cs,\ct$, in principle admit solutions with all moduli
fixed at values $\cz_0 = x_0 + i y_0$, $\cs_0 = s_0 + i\sigma_0$, and  $\ct_0 = t_0 + i \mu_0$.
We will also impose the physical requirements $\sigma_0 > 0$, $\mu_0 > 0$ and
$\im U_0 > 0$ which implies $|\Gamma| y_0 > 0$. In general existence of such solutions demands 
that the fluxes satisfy some specific properties.

In the ${\rm AdS}_4$ vacua, $\cp_2$ and $\cp_3$ are necessarily different from zero. Moreover,
combining the equations $D_\cs\cw=0$ and $\D_\ct\cw=0$ shows that at the minimum 
$\im\left(\cp_3/\cp_2\right)=0$, or equivalently
\beq
\left(\cp_3 \cp_2^* - \cp_3^* \cp_2\right)\left|_{0}\right.=0   \ .
\label{parcond}
\eeq
{}From this condition we can quickly extract useful information. For example, for the
polynomials of the nilpotent subalgebra we find that $\eps_1=0$. Similarly, for the semidirect
product $\mathfrak{su(2) \oplus u(1)^3}$, it follows that $\eps_1=-\eps_2$. Thus, in these two
cases $\cp_2$ and $\cp_3$ are forced to be parallel and equation (\ref{parcond}) is inconsequential
for the moduli. Having one equation less means that all moduli cannot be fixed. In fact, what
happens is that only a linear combination of the axions $s_0$ and $t_0$ is determined \cite{cfi}. 

Another instructive example is that of the $\mathfrak{su(2) + u(1)^3}$ subalgebra. With the polynomials
provided in table \ref{tablecps} the condition (\ref{parcond}) implies
\beq
\eps_2 - 2 \eps_1 x_0(x_0^2 + y_0^2) = 0 \ ,
\label{direx}
\eeq
where we already used that $y_0\not=0$. Now we see that forcefully $\eps_1\not=0$ because otherwise
$\eps_2$, and thus $\cp_2$ itself, would vanish. However, it could be $\eps_2=0$ and then $x_0=0$.
If $\eps_2 \not=0$ we will just have one equation that gives $y_0$ in terms of $x_0$. 

In other examples with $\cp_2$ and $\cp_3$ not parallel there are analogous results. It can happen
that (\ref{parcond}) already fixes $x_0$ or it gives $y_0$ as function of $x_0$. 
The remaining five equations can be used to obtain $\cs_0$ and $\ct_0$ in terms of $y_0$ or $x_0$, and
to find a polynomial equation that determines $y_0$ or $x_0$. This procedure can be efficiently carried out
using the algebraic package {\it Singular}. The results are described below in more detail. 

The superpotential for each $Q$-subalgebra is constructed with the flux-induced polynomials listed 
in table \ref{tablecps}. The numbers of sources needed to cancel tadpoles are given in table \ref{tabletads}.
Recall that O3-planes (D3-branes) make a positive (negative) contribution to $N_3$, whereas
O7-planes (D7-branes) yield negative (positive) values of $N_7$. 

Each supersymmetric vacua can be distinguished by the modular invariant values of the string coupling constant $g_s$ and
the potential at the minimum $V_0$ that is equal to the cosmological constant up to normalization. 
In the models at hand these quantities are given by 
\beq
V_0 =-\frac{3 |\cw_0|^2}{128\, y_0^3 \, \mu_0^3 \, \sigma_0} \qquad ; \qquad g_s = \frac1{\sigma_0} \ .
\label{vacdata}
\eeq
In all examples the vevs of the moduli $y_0$, $\sigma_0$, $\mu_0$, as well as the value $\cw_0$ of the superpotential at the minimum,
can be completely determined and will be given explicitly. It is then straightforward to evaluate the 
characteristic data $(V_0, g_s)$.

\subsubsection{Nilpotent subalgebra}
\label{sss:nilpotentres}

When $\eps_1=0$, the model based on the non-geometric fluxes of the nilpotent subalgebra is 
$U \leftrightarrow T$ dual to a IIA orientifold with only RR and NSNS fluxes already considered in the literature \cite{DeWolfe, cfi}. 
Supersymmetry actually requires $\eps_1=0$.
There are some salient features that are easily reproduced in our setup. For instance, a solution
exists only if $\lambda_3 \not=0$ and $(\lambda_1\lambda_3 - \lambda_2^2) > 0$. The axions $s_0$ and $t_0$ 
can only be fixed in the linear combination
\beq
3t_0 + \eps_2 s_0 = \frac{\lambda_2}{\lambda_3^2}(3\lambda_1\lambda_2 - 2 \lambda_2^2) \ .
\label{nilaxions}
\eeq  
The rest of the moduli are determined as
\beq
x_0=-\frac{\lambda_2}{\lambda_3} \quad ; \quad  y_0^2 = \frac{5(\lambda_1\lambda_3-\lambda_2^2)}{3\lambda_3^2}
\quad ; \quad   \sigma_0 = -\frac{2(\lambda_1\lambda_3-\lambda_2^2) y_0}{3\eps_2 \lambda_3}
\quad ; \quad \mu_0 = \eps_2 \sigma_0 \ .
\label{nilrest}
\eeq
The cosmological constant can be computed using $\cw_0=2i\mu_0 |\Gamma|^{3/2}$.

{}From the results we deduce that $\eps_2 > 0$, and $\lambda_3 > 0$ for $y_0 < 0$. Then $\im U_0 > 0$
requires $|\Gamma| < 0$ as it happens for the nilpotent algebra. The tadpole conditions then verify
$N_3 = - \lambda_3 \eps_2 |\Gamma|^3 > 0$ and $N_7 = \lambda_3 |\Gamma|^3 < 0$. The relevant conclusion
is that the model necessarily requires O3-planes and O7-planes.

\subsubsection{Semidirect sum $\mathfrak{su(2) \oplus u(1)^3}$}
\label{sss:semidirectres}

The non-geometric fluxes of this subalgebra are $U \leftrightarrow T$ dual to NSNS plus {\it geometric}
fluxes in a IIA orientifold. Models of this type have been studied previously \cite{Derendinger, vz1, cfi}.
For completeness we will briefly summarize our results that totally agree with the general solution presented 
in \cite{cfi}. Existence of a supersymmetric minimum imposes the constraint $\eps_1=-\eps_2$.
In this case it occurs again that the axions $s_0$ and $t_0$ can only be determined in a linear 
combination given by
\beq
3t_0 + \eps_2 s_0 = 3\lambda_1 + 3\lambda_2(9-7x_0) + 3\lambda_3 x_0(9-8x_0) \ .
\label{semiaxions}
\eeq  
The imaginary parts of the axiodilaton and the K\"ahler field are stabilized at values
\beq
\mu_0 = \eps_2 \sigma_0  \quad ; \quad
\eps_2 \sigma_0 = 6(\lambda_2 + \lambda_3 x_0) y_0  \ .
\label{semimusigma}
\eeq 
Notice that $\eps_2$ must be positive. It also follows that $\cw_0=2i\mu_0 (1-x_0 - iy_0)|\Gamma|^{3/2}$.
The vevs of $x_0$ and $y_0$ depend on whether the RR flux parameter $\lambda_3$ is zero or not.

When $\lambda_3=0$ we obtain
\beq
x_0=1 \quad ; \quad  3\lambda_2 y_0^2 = -(\lambda_1+\lambda_2) \ .
\label{semil3zero}
\eeq
Notice that $\lambda_2 \not=0$ to guarantee $\sigma_0 \not=0$. In fact, choosing $y_0 > 0$ it
must be $\lambda_2 > 0$. For the number of sources we find 
$N_3 = - \lambda_2 \eps_2 |\Gamma|^3 < 0$ and $N_7 = \lambda_2 |\Gamma|^3 > 0$. 
Therefore, D3 and D7-branes must be included. 

When $\lambda_3\not=0$ we instead find
\beq
\lambda_3 y_0^2 = 15(x_0-1)(\lambda_2+\lambda_3 x_0) \ ,
\label{semil3difzero}
\eeq
whereas $x_0$ must be a root of the cubic equation
\beq
160(x_0-1)^3 + 294(1+ \frac{\lambda_2}{\lambda_3})(x_0-1)^2 + 135(1+ \frac{\lambda_2}{\lambda_3})^2(x_0-1) 
+\frac{1}{\lambda_3}(\lambda_3 + 3\lambda_2 + 3\lambda_1)=0 \ . 
\label{semix0}
\eeq
The solution for $x_0$ must be real and such that $y_0 ^2 > 0$.
For the tadpoles we now have $N_7=|\Gamma|^3(\lambda_2+\lambda_3)$ and $N_3=-\eps_2 N_7$.
Thus, in general $N_3$ and $N_7$ have opposite signs. The remarkable feature is that now
they can be zero simultaneously. This occurs when the RR parameters satisfy $\lambda_2=-\lambda_3$,
in which case the cubic equation for $x_0$ can be solved exactly. 

\subsubsection{Direct sum $\mathfrak{su(2)+ u(1)^3}$}
\label{sss:directres}

As explained before, necessarily $\eps_1 \not=0$. Let us consider $\eps_2=0$  which is the condition
for $\cp_2$ to have only real roots. Now it happens that all moduli can be determined. The axions are
fixed at $x_0=0$, $s_0=0$ and $t_0=0$, whereas the real parts have vevs
\beq
y_0^2 = \frac{\eps_1\xi_3}{\xi_7} \qquad ; \qquad 
\sigma_0 = -\frac{2 \xi_7^2 y_0}{\eps_1^2 \xi_3} 
\qquad ; \qquad \mu_0 = 2 \xi_7 y_0 \ .
\label{alldprod}
\eeq 
The cosmological constant is easily found substituting $\cw_0=-2\mu_0 y_0 |\Gamma|^{3/2}$.
Clearly, the solution exists only if $\xi_3 \not=0$ and $\xi_7\not=0$. Moreover, $\eps_1\xi_3\xi_7 > 0$
and if we take $y_0 > 0$, $\xi_3 < 0$, $\xi_7 > 0$ and $\eps_1 < 0$. The numbers of sources satisfy $N_3 < 0$
and $N_7 > 0$, so that D3 and D7-branes are needed.

Taking $\eps_2\not=0$ we deduce that there are no solutions at all when $\xi_7=0$ and $\xi_3\not=0$. However, there are
minima that require $\eps_1 < 0$ and $N_7 > 0$ when $\xi_3=0$.

\subsubsection{Non-compact $\mathfrak{so(3,1)}$}
\label{sss:noncompactres}

This is the only flux configuration for which $\cp_3(\cz)$ has complex roots. It also happens that $\cp_2(\cz)$
always has three different real roots. We will briefly discuss the vacua according to whether
the NSNS flux parameter $\eps_2$ vanishes or not. 

\begin{trivlist}

\item[$\bullet$] \underline{$\eps_2=0$}

In this setup the axions are determined to be $x_0=0$, $s_0=0$ and $t_0=0$. For the imaginary parts of the
K\"ahler modulus and the axiodilaton we obtain
\beq
\mu_0 = \frac{\eps_1 \sigma_0 (3+y_0^2)}{(1-y_0^2)} \quad ; \quad 
\eps_1\sigma_0 = \frac{1}{2y_0(3+y_0^2)}\big[3\xi_7 (y_0^2-1) - \eps_1\xi_3(3y_0^2+1)\big] \ .  
\label{so31ip}
\eeq
To evaluate the potential at the minimum we use $\cw_0=2\mu_0 y_0 (1-y_0^2)|\Gamma|^{3/2}$.
Notice that $\xi_3$ and $\xi_7$ cannot be zero simultaneously and that $y_0^2=1$ is not allowed.
Actually, the imaginary part of the transformed complex structure satisfies a third order polynomial
equation in $y_0^2$ given by 
\beq
\eps_1\xi_3(5 y_0^6 + 13 y_0^4 + 15 y_0^2 -1) - \xi_7(y_0^2-1)(5 y_0^4 + 6y_0^2 -3) = 0 \ .
\label{sol31y0}
\eeq
We are interested in real roots $y_0 \not=0$ and $y_0\not=\pm1$.

Although we have not made an exhaustive analysis, it is clear that the solutions of (\ref{sol31y0}) depend
on the range of the ratio $\xi_7/\eps_1\xi_3$. For instance, there are values for which there is
no real root at all, as it occurs e.g. for  $2\xi_7=-\eps_1\xi_3$. 
For other values there might be only one real positive solution for $y_0^2$. An special example happens
when $\xi_3=0$ and the net O3/D3 charge $N_3$ is zero, while the net O7/D7 charge $N_7$ is negative as implied by the 
conditions $\mu_0 > 0$ and $|\Gamma| y_0 > 0$. Similarly,  when $\xi_7=0$ , there is only one solution in which
$N_7=0$ while $N_3 < 0$.

The third possibility is to have two allowed solutions. For instance, taking $\xi_7=2\eps_1 \xi_3$ gives roots
$y_0^2=1/5$ and $y_0^2=1+2\sqrt 2$. However, in principle the corresponding vacua cannot be realized simultaneously
because the net charges would have to jump. In fact, for $y_0^2 < 1$, it happens that $N_3 N_7 > 0$, whereas for
$y_0^2 > 1$, it must be $N_3 N_7 < 0$.  It can also arise that both solutions have $y_0^2 < 1$. 
For example, when  $\xi_7=-30\eps_1 \xi_3$ each of the two vacua has $N_3 > 0$ and $N_7 < 0$.
We will explore the phenomenon of multiple AdS vacua in more detail for the non-geometric fluxes of the
$\mathfrak{su(2)^2}$ algebra. 

\item[$\bullet$] \underline{$\eps_2\not=0$}

We have only studied the special cases when one of the flux-tadpoles $N_3$ or $N_7$ is zero.
We find that when $\eps_1=0$ the F-flat conditions can not be solved but for $\eps_1 > 0$ there are
consistent solutions for a particular range of $|\eps_2/\eps_1|$. Vacua with $\xi_3 = 0$ exist
provided that $\xi_7 < 0$. Vacua with no O7/D7 flux-tadpoles, i.e. with $\xi_7=0$, require $\xi_3 < 0$. 
One important conclusion is that for the fluxes of the non-compact $Q$-subalgebra solutions with $N_7=0$
must have $N_3 < 0$.

\end{trivlist}

\subsubsection{Compact $\mathfrak{su(2)^2}$}
\label{sss:compactres}

This is the only situation in which the polynomial $P_3(U)$ induced by the non-geometric fluxes has three
different real roots. The polynomial $P_2(U)$ generated by NSNS fluxes has complex roots
whenever $\eps_1\eps_2\not=0$, and one triple real root otherwise. We will study the vacua in both cases in some detail.

The full model based on the non-geometric fluxes of $\mathfrak{su(2)^2}$ has an interesting residual symmetry
that exchanges the  NSNS auxiliary parameters. It can be shown that the effective action is invariant under 
$\eps_1 \leftrightarrow \eps_2$,  $\xi_3 \to \xi_3$ and $\xi_7 \to \xi_7$, together with the field transformations 
\beq
\cz \to 1/\cz^* \qquad ; \qquad  \cs \to -\cs^* \qquad ; \qquad \ct \to -\ct^* \ . 
\label{extrasym}
\eeq
This symmetry leaves one of the $\cp_3$ roots invariant while exchanging the other two.

\subsubsection*{\thesubsubsection.1 \quad $P_2(U)$ with triple real root}
\label{su2zero}
\addcontentsline{toc}{subsubsection}{\hspace{13pt} \thesubsubsection.1 \quad $P_2(U)$ with triple real root }

Due to the symmetry (\ref{extrasym}) it is enough to consider $\eps_1=0$ and $\eps_2 \not=0$.
In this model the axions are stabilized at vevs
\beq
x_0=-\frac12 \qquad ; \qquad \eps_2s_0=3\xi_7 -\frac{\eps_2\xi_3}{2}  \qquad ; \qquad t_0=\xi_7 -\frac{\eps_2\xi_3}{2} \ . 
\label{su2rst}
\eeq
The imaginary parts of the K\"ahler modulus and the axiodilaton are fixed in terms of $y_0$ according to 
\beq
\mu_0 = -\frac{4\eps_2 \sigma_0}{(1+4y_0^2)} \quad ; \quad 
\eps_2\sigma_0 = -y_0\big[3\xi_7  + \frac{\eps_2 \xi_3}{8}(4y_0^2-3)\big] \ .
\label{su2ip}
\eeq
At the minimum $\cw_0=2i \eps_2 \sigma_0 |\Gamma|^{3/2}$.
Clearly $\xi_3$ and $\xi_7$ cannot vanish simultaneously so that the model always requires additional
sources to cancel tadpoles. Observe that necessarily $\eps_2 < 0$.

The modulus $y_0$ is determined by the fourth order polynomial equation  
\beq
\eps_2\xi_3(4y_0^2-1)(4y_0^2+5) - 8\xi_7(4y_0^2-5)= 0 \ .
\label{sol2y0}
\eeq
In the two special cases $\xi_7=0$ and $\xi_3=0$ an exact solution is easily found. When $\xi_3 \xi_7 \not=0$ there
can be two AdS solutions. The corresponding vacua, which can be characterized by the net tadpoles $N_3$ and $N_7$, are
described more extensively in the following.

\begin{trivlist}

\item[$\bullet$] \underline{$N_7=0$}

When $\xi_7=0$ the vevs have the very simple expressions
\beq
y_0^2 = \frac14
\qquad ; \qquad   \sigma_0 = \frac{\xi_3 y_0}{4}
\qquad ; \qquad \mu_0 = -2\eps_2 \sigma_0 \qquad ; \qquad
V_0 = \frac{12 |\Gamma|^3 y_0}{\eps_2 \xi_3^2}
\ .
\label{n7zero}
\eeq 
Since both $\mu_0$ and $\sigma_0$ are positive, it must be $\eps_2 <0$, and taking $y_0 > 0$, $\xi_3 > 0$.
Therefore, $N_3 > 0$ and O3-planes must be included.

\item[$\bullet$] \underline{$N_3=0$}

This is the case $\xi_3=0$. The moduli and the cosmological constant are fixed at values
\beq
y_0^2 = \frac54
\qquad ; \qquad   \eps_2\sigma_0 = -3\xi_7 y_0
\qquad ; \qquad \mu_0 = -\frac23 \eps_2 \sigma_0  \qquad ; \qquad
V_0 = \frac{9 |\Gamma|^3 \eps_2 y_0}{500\, \xi_7^2}
\ .
\label{n3zero}
\eeq 
Necessarily $\eps_2 < 0$, and choosing  $y_0 > 0$, $\xi_7 > 0$.
Hence, $N_7 > 0$ and D7-branes are required.

\item[$\bullet$] \underline{$N_3N_7\not=0$}

The solutions for $y_0$ depend on the ratio $\xi_7/\eps_2\xi_3$. A detailed analysis can be easily performed because
the polynomial equation (\ref{sol2y0}) is quadratic in $y_0^2$. We find that there are no real solutions in the interval 
$\msm{1/8 < \xi_7/\eps_2\xi_3 < (7+2\sqrt{10})/4}$. On the other hand, when $\msm{0 < \xi_7/\eps_2\xi_3 < 1/8}$, there is only one real positive
solution for $y_0^2$ and it requires $N_3 > 0$ and $N_7 < 0$.   
For $\msm{\xi_7/\eps_2\xi_3 \leq 0}$ there is only one acceptable root for $y_0^2$
and it leads to $N_3 > 0$ and $N_7 \geq 0$. A more interesting range of parameters is $\msm{\xi_7/\eps_2\xi_3 > (7+2\sqrt{10})/4}$ because
there are two allowed solutions for $y_0^2$ and for both it must be that $N_3 < 0$ and $N_7 > 0$. The upshot is that there can be metastable
AdS vacua in the presence of D3 and D7-branes.  

\end{trivlist}

\subsubsection*{\thesubsubsection.2 \quad $P_2(U)$ with complex roots}
\label{su2nonzero}
\addcontentsline{toc}{subsubsection}{\hspace{13pt} \thesubsubsection.2 \quad $P_2(U)$ with complex roots}

The F-flat conditions can be unfolded to obtain analytic expressions for the vevs of all moduli.
However, for generic range of parameters, a higher order polynomial equation has to be solved to determine $y_0$ in the end.
The main interesting feature is the appearance of multiple vacua even when $N_3 N_7 = 0$, i.e. when there are either no O7/D7 or no O3/D3 net charges 
present. We will first describe the overall picture and then present examples. For definiteness we always choose $y_0 > 0$ so that
$|\Gamma| > 0$ is required to have $\im U_0 > 0$ for the complex structure.

To obtain and examine the results it is useful to make some redefinitions. The idea is to leave as few free
parameters as possible in the F-flat equations. 
Since $\epsilon_1$ is different from zero we can work with the ratio
\beq
\rho=\frac{\epsilon_2}{\epsilon_1} \ .
\label{newps}
\eeq
By virtue of the residual symmetry (\ref{extrasym}) there is an invariance under $\rho \to 1/\rho$. Therefore, we can restrict 
to the range $-1 \leq \rho \leq 1$, where the boundary corresponds to the fixed points of the inversion. Furthermore, as discussed at the 
end of section \ref{subsubso4}, the parameter $\rho$ is either a rational number or involves at most square roots of rationals.

When $\xi_3 \not=0$ it is also convenient to introduce new variables as
\beq
\ct=\epsilon_1 \xi_3 \,\hat\ct
\qquad ; \qquad
\cs = \xi_3 \hat \cs \qquad ; \qquad \xi_7 = \eps_1 \xi_3 (\rho^2+1) \eta \ .
\label{newus}
\eeq
The definition of the parameter $\eta$ seems awkward but it simplifies the results. Notice that $\eta \to \eta\rho$ under (\ref{extrasym}).
In the new variables the superpotential becomes
\beq
\cw=|\Gamma|^{3/2} \epsilon_1 \xi_3 [3\, \hat\ct \cz(\cz+1) + \hat\cs(\cz^3 + \rho) + (1-\rho \cz^3) + 3\eta (1+\rho^2) \cz (1-\cz) ] \ .
\label{w1}
\eeq
Since the F-flat conditions are homogeneous in $\cw$ the resulting equations will only depend on the parameters $\rho$ and $\eta$.
When $\xi_3=0$ we just make different field redefinitions, i.e. $\ct=\epsilon_1 \xi_7 \,\hat\ct$ and $\cs=\xi_7 \,\hat\cs$, so that   
the free parameters will be $\rho$ and $\xi_7/\eps_1$.

Manipulating the F-flat conditions enables us to find the vevs $\ct_0$ and $\cs_0$ as functions of $(x_0, y_0)$. 
The expressions are tractable but bulky so that we refrain from presenting them. The exception is the 
handy relation between the size and string coupling moduli 
\beq
\mu_0=\frac{\eps_1 \sigma_0(3x_0^2-y_0^2)}{1+2x_0} \ ,
\label{msvacgen}
\eeq
which is valid when $x_0\not=\ds{\msm{-\frac12}}$ and $y_0^2\not=\ds{\msm{\frac34}}$. There is a solution with  $x_0 =\ds{\msm{-\frac12}}$ and 
$y_0^2 =\ds{\msm{\frac34}}$ but it has $\mu_0=-\eps_1(1+\rho)\sigma_0$, \, $\mu_0=3\xi_7y_0$, and it requires $\eta=-(1+\rho)/(\rho^2-7\rho+1)$. There is another
vacuum with  $x_0=\ds{\msm{-\frac12}}$ that occurs when $\rho \to \infty$ ($\eps_1=0$) and was discussed in section \ref{su2zero}. 
The case $x_0^2=y_0^2$, which is better treated separately, requires $\xi_7\not=0$ unless $\rho=0$.

The residual unknowns $(x_0, y_0)$ are determined from the coupled system
\beqa
&{\hspace*{-8mm}} & y_0^4+2x_0(1+x_0)y_0^2-\rho(2x_0+1) + x_0^3(x_0+2) = 0 \label{eqab} \ , \\[4mm]
& {\hspace*{-8mm}} & 
y_0^6 \left( 1+2\eta x_0-2 \eta \right) + \left( 1+30 \eta {x_0}^{3}-{x_0}^{2}+18\eta {x_0}^{2}-6\,\rho\eta \right) y_0^4
\nonumber \\
& {\hspace*{-8mm}} & 
+ \, x_0 \left( 54\eta {x_0}^{4}+11 {x_0}^{3}+42 \eta {x_0}^{3}+8 {x_0}^{2}+12 \rho \eta x_0-4 x_0-6 \rho\eta \right) y_0^2
\label{eqab2} \\
& {\hspace*{-8mm}} & 
\mbox{}+ \left( 2\rho+4\rho x_0+11 {x_0}^{3}+13 {x_0}^{4} \right)  \left( 2\,\rho\eta+2\eta{x_0}^{3}+{x_0}^{2}+x_0 \right) = 0
\ . \nonumber
\eeqa
The corresponding equations when $\xi_3=0$ can be obtained taking the limit $\eta \to \infty$.
Eliminating $y_0$ for generic parameters gives a ninth-order polynomial equation for $x_0$.

For some range of parameters the above equations can admit several solutions for $\cz_0=x_0+iy_0$, which in turn yield consistent
values for the remaining moduli. The existence of multiple vacua is most easily detected in the limiting cases in which one of the net tadpoles $N_7$
or $N_3$ vanishes, equivalently when $\xi_7=0$ ($\eta=0$) or  $\xi_3=0$ ($\eta \to \infty$). In either limit the NSNS
parameter $\rho$ can still be adjusted. We expect the results to be invariant under $\rho \to 1/\rho$
and this is indeed what happens.

We have mostly looked at models having no O7/D7 net charge, namely with $\eta=0$.
It turns out that the solutions require $\xi_3 > 0$ so that $N_3 > 0$ and O3-planes must be present.
Below we list the main results.

\noindent
1. For $\rho=1$ there are no minima with moduli stabilized.

\noindent
2.  For $\rho=-1$ there is only one distinct vacuum with data
\beq
\mfn{
\cz_0 =    -0.876 + 1.158 \, i  \quad ; \quad
\cs_0  =    \xi_3(-0.381 + 0.238\, i) \quad ; \quad 
\ct_0  =   \epsilon_1 \xi_3 (0.602 -  0.305\, i) \quad  ; \quad
V_0= \frac{2.38\, |\Gamma|^3}{\xi_3^2 \epsilon_1}
\ . } 
\label{solrmu}
\eeq
Notice that necessarily $\xi_3 > 0$ and $\epsilon_1 < 0$. Actually, for $\rho=-1$, there is a second consistent
solution but it is related to the above by the residual symmetry (\ref{extrasym}).

\noindent 
3. There can be only one solution when $\rho_c \leq \rho < 1$, where $\rho_c=-0.7267361874$.
The critical value $\rho_c$ is such that the discriminant of the polynomial equation that determines
$x_0$ is zero. Consistency requires $\eps_1 < 0$ and $\xi_3 > 0$ so that O3-planes are needed.
For instance, when $\rho=0$ the solution is exact and has
\beq
\mfn{
\cz_0 =  -1 + i \quad ; \quad 
\cs_0 =  \ds{\frac{\xi_3}8}(4 + i) \quad ; \quad
\ct_0 = \ds{\frac{\epsilon_1 \xi_3}4}(2-i) \quad ; \quad
V_0= \frac{6\, |\Gamma|^3}{\xi_3^2 \epsilon_1} 
\ .}
\label{solrzero}
\eeq
As expected, upon the transformation (\ref{extrasym}) this vacuum coincides with that having $\xi_7=0$ and $\eps_1=0$, given in (\ref{n7zero}).
For other values of $\rho$ the solution is numerical. For example, taking $\rho=\msm{\frac12}$ leads to the vevs
\beq
\mfn{
\cz_0=-1.036 + 0.834\, i  \quad ; \quad \cs_0=\xi_3(1.561 + 0.192) \quad ; \quad
\ct_0=\xi_3 \epsilon_1(1.055 - 0.453\, i) \quad ; \quad V_0=\frac{2.283 |\Gamma|^3}{ \xi_3^2 \, \epsilon_1} \ .} 
\label{solrm12}
\eeq

\noindent
4. The important upshot is that in the interval $-1 < \rho < \rho_c$ there can be two 
distinct solutions for the same set of fluxes. An example with $\rho=\msm{-\frac45}$ is shown in table \ref{solezero}.
Notice that the last two solutions can exist for $\xi_3 > 0$ and
$\epsilon_1 > 0$. The first solution can also occur but for $\xi_3 > 0$ and
$\epsilon_1 < 0$. 

\begin{table}[htb]
\renewcommand{\arraystretch}{1.15}
\begin{center}
{\footnotesize
\begin{tabular}{|c|c|c|c|}
\hline
$\cz_0$ & $\cs_0/\xi_3$ & $\ct_0/\xi_3 \epsilon_1$ & $V_0\, \xi_3^2 \, \epsilon_1/|\Gamma|^3$
\\
\hline
$\!\! -0.91105442 + 1.14050441 \, i$ &  $ -0.26002362 + 0.19059447 \, i$ &     
$0.53128071 - 0.27572497\, i$ & 3.353 \\
\hline
$\!\! -0.43550654 + 0.73478523  \, i$ & $ 0.28605555 +  0.55017649 \, i$ &    
$0.60410811 + 0.12407321 \, i$ &  \!\!\! -2.168    \\
\hline
$\!\! -0.40368586 + 0.57866160\, i$ & $ 0.49215445 + 0.33255331 \, i$ &    
$0.57101568 + 0.26593032\, i$ &  \!\!\! -1.880  \\
\hline
\end{tabular}
}
\end{center}
\caption{\small Degenerate vacua for $\xi_7=0$ and $\rho=\msm{-\frac45}$.}
\label{solezero}
\end{table}
\noindent

For models having no O3/D3 net charge a detailed analysis is clearly feasible but we have only sampled narrow ranges of the adjustable parameter $\rho$. 
Consistent solutions must have $\eps_1 < 0$ and $\xi_7 > 0$. Hence, $N_7 > 0$ and D7-branes must be included. There are values of $\rho$, e.g. $\rho=-1$, 
for which there are no vacua with stabilized moduli. For $\rho=1$ there is only one minimum which can be computed exactly.
More interestingly, models of this type can also exhibit multiple vacua. In table \ref{soleinfty} we show one example with $\rho=\msm{\frac34}$.
Observe that both solutions exist for $ \eps_1 < 0$ and $\xi_7 >0$.
\begin{table}[htb]
\renewcommand{\arraystretch}{1.15}
\begin{center}
{\footnotesize
\begin{tabular}{|c|c|c|c|}
\hline
$\cz_0$ & $\eps_1\cs_0/\xi_7$ & $\ct_0/\xi_7$ & $V_0\, \xi_7^2 /\epsilon_1|\Gamma|^3$
\\
\hline
$-0.88312113 + 0.74580943 \, i$ &  $  -6.1818994 - 1.6867660 \, i$ & 
$-4.20643209 +  3.92605399 \, i$ & 0.026 \\
\hline
$0.20646056  +  0.89488895 \, i$ & $  0.03039439 - 2.49813344  \, i$ &  
$-0.06455485 +  1.18981502 \, i$ &     0.084 \\
\hline
\end{tabular}
}
\end{center}
\caption{\small Vacua for $\xi_3=0$ and $\rho=\msm{\frac34}$.}
\label{soleinfty}
\end{table}

\section{Aspects of the non-geometric landscape}
\label{sec:lands}

In this section we discuss the main aspects of the ${\rm AdS}_4$ vacua in our models that are standard examples of type IIB toroidal 
orientifolds with O3/O7-planes. Besides the axiodilaton $S$, after an isotropic Ansatz the massless scalars reduce to the overall complex 
structure $U$ and the size modulus $T$. Fluxes of the RR and NSNS 3-forms generate a potential that gives masses only to $S$ and $U$. 
The new ingredient here are non-geometric $Q$-fluxes, that are required to restore T-duality between type IIA and type IIB, and that
induce a superpotential for the K\"ahler field $T$. The various fluxes must satisfy certain constraints arising from Jacobi or Bianchi identities.
The problem is then to minimize the scalar potential while solving the constraints. The question is whether there are solutions
with all moduli stabilized. We have seen that the answer is affirmative and now we intend to analyze it in more detail. 

It is instructive to begin by recounting the findings of the previous sections. The initial step is to classify the subalgebras whose
structure constants are the $Q$'s. With the isotropic Ansatz there are only five classes. For each type, the non-geometric fluxes can be written in terms
of four auxiliary parameters $\genfrac{(}{)}{0pt}{}{\a \, \b}{\g \, \d}= \Gamma$, in such a way that the Jacobi identities are automatically satisfied.  
Other fluxes can also be parametrized using  $\Gamma$ plus additional variables: $(\eps_1, \eps_2)$ for NSNS, and $(\xi_3, \xi_7, \xi_s, \xi_t)$
or $(\lambda_1, \lambda_2, \lambda_3, \xi_s, \xi_t)$ for RR. The significance of $\Gamma$ is that it defines a transformed complex 
structure $\cz=(\a U + \b)/(\g U + \d)$ that is invariant under the modular group $SL(2,\Z)_U$. The effective action can be expressed in
terms of $\cz$ according to the $Q$-subalgebra. Once the subalgebra is chosen the vacua will depend only on the variables 
$\Gamma$, $(\eps_1, \eps_2)$, and $(\xi_3, \xi_7)$ or $(\lambda_1, \lambda_2, \lambda_3)$, that in turn determine the values of the
cosmological constant and the string coupling $(V_0, g_s)$, as well as the net tadpoles $(N_3, N_7)$. In many examples, the vevs of the moduli
can be determined in closed form.

Our approach to analyze the vacua in presence of non-geometric fluxes has the great advantage that the degeneracy due to modular transformations
of the complex structure is already taken into account. Inequivalent vacua are just labelled by the vevs $(\cz_0, S_0, T_0)$ that are
modular invariant. In practice this means that we can study families of modular invariant vacua by choosing a particular structure for
$\Gamma$. In section \ref{sub:fam} we will give concrete examples.

There is an additional vacuum degeneracy because the characteristic data $(V_0, g_s)$ happen to be independent of the parameters $(\xi_s, \xi_t)$. 
The explanation is that they correspond to shifts of the axions $\re S$ and $\re T$ which can be reabsorbed in the RR fluxes.
The flux-induced RR tadpoles $(N_3, N_7)$ are blind to $(\xi_s, \xi_t)$ as well.
Apparently, generic shifts in $\re S$ and $\re T$ are not symmetries of the compactification, so that 
two vacua differing only in the RR flux parameters  $(\xi_s, \xi_t)$ would be truly distinct. We argue below that the vacua are
equivalent because the full background is symmetric under $S \to S - \xi_s$, and $T \to T - \xi_t$. 

In absence of non-geometric fluxes the 3-form RR field strength that appears in the 10-dimensional action 
is given by $F_3=dC_2- H_3 \wedge C_0 + \bar F_3$, where $H_3 = dB_2 + \bar H_3$. The natural generalization to include
non-geometric fluxes is
\beq
F_3=dC_2- H_3 \wedge C_0 + QC_4 + \bar F_3 \ ,
\label{f3q}
\eeq  
where $QC_4$ is a 3-form that we can extract from (\ref{QJexpan}) because $\re \cj = C_4$. In fact, $C_4=-\re T \sum_I \tilde \omega^I$, where
$\tilde \omega ^I$ are the basis 4-forms. Recall also that $C_0 = \re S$. Notice then that $F_3$ involves the axions in question.
The relevant result is that $F_3$ is invariant\footnote{We thank P. C\'amara for giving us this hint.}
under the shifts $S \to S - \xi_s$, and $T \to T - \xi_t$. To show this
we first compute the variation of $\bar F_3$ using the universal terms (\ref{uniRR}) in the parametrization of the RR fluxes and
then substitute in (\ref{f3q}). In the effective \deq4 action the result is simply that the superpotential is invariant
under these axionic shifts and the corresponding transformation of the RR fluxes. In turn this follows from (\ref{P3Iso}) after 
substituting (\ref{uniRR}).

\subsection{Overview}
\label{ss:view}

We now describe in order some prominent features of the ${\rm AdS}_4$ vacua with non-geometric $Q$-fluxes switched on. 

\bigskip
\noindent
1. The explicit results of section \ref{sub:ads} indicate that in all models the vevs $\sigma_0 = \im S_0$ and $\mu_0 = \im T_0$ are
correlated. This generic property follows from the F-flat conditions simply because the superpotential is linear in the axiodilaton
and the K\"ahler modulus. Recall that the vevs in question determine physically important quantities, namely the string coupling $g_s=1/\sigma_0$,
and the overall internal volume $V_{int}=(\mu_0/\sigma_0)^{3/2}$. To trust the perturbative string approximation $g_s$ must be small and we  
will shortly explain, as already shown in  \cite{stw2}, that generically there are regions in flux space in which both $g_s$ 
and the cosmological constant are small, while $V_{int}$ is large. We stress again the caveat that even at large overall volume there could still 
exist light winding string states when non-geometric fluxes are in play. These effects are certainly important in trying to lift the solutions
to full string vacua. In this paper we only claim to have found vacua of the effective field theory with a precise set of massless fields
and interactions due to generalized fluxes.

\bigskip
\noindent
2. Another common feature of all models is the relation between moduli vevs and net RR charges. In type IIB toroidal orientifolds
it is known that in Minkowski supersymmetric vacua the contribution of RR and NSNS fluxes to the $C_4$ tadpole is positive ($N_3 >0$)
and this occurs if and only if $\im S_0 > 0$ \cite{kst}. The interpretation is that to cancel the tadpole due to $\bar F_3$
and $\bar H_3$ it is mandatory to include O3-planes, whereas D3-branes can be added only as long as $N_3$ stays positive.
This is also true for no-scale Minkowski vacua in which supersymmetry is broken by the F-term of the K\"ahler field.
Turning on non-geometric fluxes enables to stabilize all moduli at a supersymmetric ${\rm AdS}_4$ minimum. At the same time, 
the $Q$-fluxes induce a $C_8$ tadpole of magnitude $N_7$ that can be cancelled by adding O7-planes and/or D7-branes.
We find in general that the vevs $\im S_0$ and $\im T_0$, that must be positive, are correlated to the tadpoles $(N_3, N_7)$.
According to the $Q$-subalgebra there are several possibilities for the type of sources that have to be included. 
For example, the models considered in \cite{stw2}, having  $N_3 > 0$ and $N_7 =0$, proceed only with the fluxes of
the compact $\mathfrak{su(2)^2}$. 

\medskip
\noindent
For the $Q$-fluxes of the nilalgebra, and the semidirect sum $\mathfrak{su(2) \oplus u(1)^3}$, there is a relation $N_3 = -\eps_2 N_7$,
with $\eps_2 > 0$. Only in the latter case it is allowed to have $N_3=N_7=0$, and the sources can be avoided altogether.
For the fluxes of $\mathfrak{su(2) + u(1)^3}$ it turns out that orientifold planes are unnecessary to cancel tadpoles, but both D3 and D7-branes
must be added ($N_3 < 0$, $N_7 > 0$). 

\medskip
\noindent
The fluxes of the semisimple subalgebras are more flexible. In particular, it can happen that
one flux-tadpole vanishes while the other must have a definite sign. Moreover, the sign is opposite for the compact and non-compact
cases. For instance, when $N_7=0$, $N_3 > 0$ and O3-planes are obligatory for the $\mathfrak{su(2)^2}$ fluxes, while for $\mathfrak{so(3,1)}$
$N_3 < 0$ and D3-branes are required. 

\medskip
\noindent
The magnitudes of the vevs are also proportional to the net tadpoles. This then implies that the string coupling typically decreases
when $N_3$ and/or $N_7$ increase. However, the number of D-branes cannot be increased arbitrarily without taking into account their
backreaction.

\bigskip
\noindent
3. Consistency of the vacua can in fact be related to the full 12-dimensional algebra in which
the $\bar H$ and $Q$-fluxes are the structure constants. The reason is that the conditions $\im S_0 > 0$ and $\im T_0 > 0$ also impose restrictions on the
signs of the NSNS parameters $(\eps_1, \eps_2)$. For instance, in section \ref{sss:compactres} we have seen that for $Q$-fluxes of the
compact $\mathfrak{so(4) \sim su(2)^2}$, the solutions with $\eps_1=0$ require $\eps_2 < 0$. This in turn implies, as explained in section
\ref{subsubso4}, that the full gauge algebra is $\mathfrak{so(4) + iso(3)}$. Another simple example is the model based on the  
$\mathfrak{su(2)+ u(1)^3}$ $Q$-subalgebra. The vacua of \ref{sss:directres} with $\eps_2=0$ require $\eps_1 < 0$ and it can then be shown
that the full gauge algebra is  $\mathfrak{so(4) + u(1)^6}$.
A more detailed study of the 12-dimensional algebras is left for future work \cite{guarino}.

\bigskip
\noindent
4. We defer to section \ref{sub:fam} a more thorough discussion of the landscape of values attained by the string coupling $g_s$ 
and the cosmological constant $V_0$, for the fluxes of the compact $\mathfrak{su(2)^2}$ $Q$-subalgebra. The situation for $\mathfrak{so(3,1)}$
is similar and can be analyzed using the results of section \ref{sss:noncompactres}. The model based on the direct product  $\mathfrak{su(2)+u(1)^3}$
is different because both $N_3$ and $N_7$ must be non-zero, but it can still be shown that there exist vacua with small $g_s$ and $V_0$.
The models built using the nilpotent and semidirect $Q$-subalgebras have been studied in their T-dual IIA formulation in refs. \cite{DeWolfe, cfi},
where it was found that there are infinite families of vacua within the perturbative region.

\bigskip
\noindent
5. A peculiar result is the appearance of multiple vacua for certain combination of fluxes. These events occur only in models based
on the semisimple $Q$-subalgebras. They can have $N_3 N_7=0$ or $N_3 N_7 \not=0$, but in the former case both NSNS parameters 
$(\eps_1, \eps_2)$ must be non-zero. Reaching small string coupling and cosmological constant typically requires that $N_3$ and/or $N_7$
be sufficiently large.

\bigskip
\noindent
6. To cancel RR tadpoles it might be necessary to add stacks of D3 and/or D7-branes. These additional D-branes could also
generate a charged chiral spectrum but more generally a different sector of D-branes will serve this purpose. In any case,
the D-branes that can be included are constrained by cancellation of Freed-Witten anomalies \cite{cfi, vz2}.   
In absence of non-geometric fluxes the condition amounts to the vanishing of $\bar H_3$ when integrated over any internal 3-cycle wrapped by
the D-branes. For unmagnetized D7-branes in $\T^6/\Z_2 \times \Z_2$, with $\bar H_3$ given in (\ref{H3expan}), it is easy to see that the condition is met, 
whereas for D3-branes it is trivial. When $Q$-fluxes are switched on the modified condition \cite{vz2} is still satisfied basically because the 3-form
$Q\cj$, defined in (\ref{QJexpan}), can be expanded in the same basis as $\bar H_3$.

\medskip
\noindent
D3-branes and unmagnetized D7-branes in $\T^6/\Z_2 \times \Z_2$ do not give rise to charged chiral matter. Therefore the models
will not have $U(1)$ chiral anomalies. This is consistent with the fact that the axions $\re S$ and $\re T$ are generically stabilized
by the fluxes and having acquired a mass they could not participate in the Green-Schwarz mechanism to cancel the chiral 
anomalies\footnote{We thank L. Ib\'a\~nez for discussions on this point.}.

\medskip
\noindent
To construct a more phenomenologically viable scenario one could introduce magnetized D9-branes as in the $\T^6/\Z_2 \times \Z_2$
type IIB orientifolds with NSNS and RR fluxes that were considered some time ago \cite{magnetized}. 
Now, care has to be taken because magnetized D9-branes suffer from Freed-Witten anomalies. They are actually forbidden in absence
of non-geometric fluxes when $\bar H_3 \not=0$. 

\medskip
\noindent
The effect of the $Q$-fluxes can be studied as explained in \cite{vz2}. Cancellation of Freed-Witten anomalies
translates into invariance of the superpotential under shifts \mbox{$S \to S + q_s \nu$} and   
\mbox{$T \to T + q_t \nu$}, where the real charges $(q_s,q_t)$ depend on the $U(1)$ gauged by the D-brane.
Applying this prescription we conclude that in our setup with isotropic fluxes magnetized D9-branes could be introduced only in
models based on the nilpotent and semidirect sum $\mathfrak{su(2) \oplus u(1)^3}$ \mbox{$Q$-subalgebras}. The reason is that only in these
cases the flux-induced polynomials $P_2(U)$ and $P_3(U)$ can be chosen parallel and then $W$ can remain invariant under the
axionic shifts. Equivalently, only in these cases the axions $\re S$ and $\re T$ are not fully determined and the residual massless
linear combination can give mass to an anomalous $U(1)$. For other \mbox{$Q$-subalgebras} the polynomials  
$P_2(U)$ and $P_3(U)$ are linearly independent and both axions are completely stabilized. 

\medskip
\noindent
It would be interesting to study the consistency conditions on magnetized D9-branes in models with non-isotropic fluxes. In principle
there could exist configuration of fluxes such that the general superpotential (\ref{fullW}-\ref{p3gen}) is invariant under axionic
shifts of $S$ and the K\"ahler moduli $T_I$. 

\subsection{Families of modular invariant vacua}
\label{sub:fam}

To generate specific families of vacua we first choose the $Q$-subalgebra and then select the parameters in $\Gamma$. 
In general $\Gamma$ can be chosen so that the non-geometric fluxes are even integers. The NSNS fluxes turn out to be even integers 
by picking $(\eps_1, \eps_2)$ appropriately. One can also start from given non-geometric and NSNS 
even integer fluxes and deduce the corresponding $\Gamma$ and  $(\eps_1, \eps_2)$.  Similar remarks apply to the RR fluxes.   
We will illustrate the procedure for the compact $\mathfrak{su(2)^2}$. 

If one of the parameters vanishes, say $\g=0$, it can be shown from (\ref{LimC}) that the ratios $\d/\a$ and $\b/\a$ are rational numbers
(recall that $|\Gamma|\not=0$ so that $\a, \d \not=0$). It then follows that by a modular transformation, c.f. (\ref{Gmodt}), we can go to
a canonical gauge in which also $\b=0$. 

The canonical diagonal gauge $\g=\b=0$ is completely generic when $\eps_2=0$ ($\eps_1 \not=0$). In this case we find that $\b/\a$ and $\g/\d$
are rational because they are given respectively by quotients of NSNS and non-geometric fluxes. Therefore, $\b$ and $\g$ can be gauged away
by modular transformations. If instead $\eps_1=0$, but $\eps_2\not=0$, we can take $\a=\d=0$.

When $\eps_1 \eps_2 \not=0$ we can still use the canonical gauge but it will not give the most general results that are obtained
simply by considering $\a, \b, \g, \d \not= 0$.

\subsubsection{Canonical families for $\mathfrak{su(2)^2}$ fluxes}
\label{ss:canonical}

For each subalgebra we can obtain families of vacua starting from the canonical gauge defined by $\g=\b=0$. 
In the $\mathfrak{su(2)^2}$ case only the non-geometric fluxes $\tilde c_1$ and $\tilde c_2$ are 
different from zero and can be written as
\beq
\tilde c_1 = -2m \quad ; \quad \tilde c_2 = 2n  \quad ; \quad m, \, n \, \in \Z  \ .
\label{nongeocan}
\eeq 
{}From (\ref{LimC}) we easily find $\a/\d=n/m$, $\d^3=2m^2/n$, so that
$|\Gamma|^3=4 nm$. The non-zero NSNS and RR fluxes are easily found to be 
\beqa
b_0 & = & - \frac{2m^2}{n}\eps_2 \quad ; \quad  b_3 = \frac{2n^2}{m}\eps_1 \quad ; \quad 
a_0  =  \frac{2m^2}{n}(\eps_1 \xi_3 + \eps_2 \xi_s) 
\ , \label{abcan} \\ 
a_1 & = &  -2m(\xi_t + \xi_7) \quad ; \quad
a_2 = 2n(\xi_t - \xi_7) \quad ; \quad a_3 = -\frac{2n^2}{m}(\eps_1\xi_s - \eps_2 \xi_3)  \ .
\nonumber
\eeqa
Since the $b$'s and $a$'s are (even) integers, it is obvious that $(\eps_1, \eps_2)$ and $(\xi_3, \xi_7, \xi_s, \xi_t)$ are all
rational numbers.

The moduli vevs depend on $(\xi_3, \xi_7)$ and $(\eps_1, \eps_2)$. For concreteness, and to compare with the results of \cite{stw2},
we focus on the case $\xi_7=0$. Other cases can be studied using the results of section \ref{sss:compactres}. 
When $\xi_7=0$ the RR fluxes $a_1$ and $a_2$ are spurious, they can be eliminated by setting $\xi_t=0$, 
i.e. by a shift in  $\re T$. 

To continue we have to distinguish whether one of the NSNS parameters $\eps_1$ or $\eps_2$ is zero. Recall that in this case
the flux induced polynomial $P_2$ does not have complex roots. 

\begin{trivlist}

\item[$\bullet$] \underline{$\eps_1 \eps_2=0$}
 
Let us consider $\eps_2=0$. Then, also $a_3$, or $\xi_s$, is irrelevant and can be set to zero by a shift in $\re S$. 
The important physical parameters are $\eps_1$ and $\xi_3$, they can be deduced from $b_3$ and $a_0$.  
Notice also that at this point $N_3=a_0 b_3$. 
Using (\ref{solrzero}) we obtain the values of the cosmological constant and the string coupling
\beq
V_0 = \frac{48 \, m^6 b_3^3}{n^3 N_3^2} \qquad ; \qquad g_s = \frac{8\, m^3 b_3^2}{n^3 N_3} \ .
\label{candata}
\eeq
Consistency requires $\eps_1 < 0$ and $\xi_3  > 0$, or equivalently $V_0 < 0 $ and $g_s > 0$.
For the purpose of counting distinct vacua we can safely assume $b_3 > 0$ and then $m, n < 0$. 

As noticed in \cite{stw2}, the important outcome is that $g_s$ and $V_0$ can be made arbitrarily small by keeping
$b_3$ and $m$ fixed while letting $n \to \infty$. 

In our approach it is also easy to see that $(V_0, g_s)$ always take values of the form (\ref{candata}) whenever $P_2$ has only real roots.
This follows because all vacua are related by modular transformations plus axionic shifts. 
However, if as in \cite{stw2} we want to count the vacua with fluxes bounded by an upper limit $L$, it does not suffice to just consider
the canonical gauge. The reason is that by performing modular transformations and axionic shifts we can reach larger effective
values of $b_3$ that seem to violate the tadpole condition. Rather than an elaborate argument we will just provide a simple example.
We can go to a non-canonical gauge with $\g=0$ but $\b\not=0$ and also take $\xi_t=0$ but $\xi_s \not=0$. With these choices it is
straightforward to show that $N_3=a_0 b_3 - a_3 b_0$, which would allow to take e.g. $b_3=N_3$ that is forbidden when $b_0=0$ ($\b=0$),
or $a_3=0$ ($\xi_s=0$), because $a_0$ must be even. To do detailed vacua statistics it is necessary to use generic gauge and axionic shifts.

\item[$\bullet$] \underline{$\eps_1 \eps_2\not=0$}

As in section \ref{su2nonzero} we set $\eps_2=\rho \eps_1$. In the canonical gauge the parameter $\rho$ is a rational number that we assume to
be given. We choose to vary the NSNS flux $b_3$ that determines
\beq
\eps_1= \frac{m b_3}{2n^2} \quad ; \quad b_0 = -\frac{\rho \, m^3 b_3}{n^3} \ , 
\label{bzero}
\eeq
where $m, n$ are the integers coming from the non-geometric fluxes. The vacuum data have been found to be
\beq
V_0 = \frac{4 F_V \, n m }{\eps_1 \xi_3^2} \qquad ; \qquad g_s = \frac{1}{F_g \xi_3} \ ,
\label{candatagen}
\eeq
where we used $|\Gamma|^3=4nm$. The numerical factors $F_V$ and $F_g$ depend on $\rho$. For instance, for $\rho=0$, $F_V=6$ and $F_g=\msm{1/8}$.
Other examples are given in section \ref{su2nonzero}. We remark that for $\rho$ in a particular range there can be multiple vacua, meaning
that for some $\rho$ the above numerical factors might take different values (e.g. table \ref{solezero}).    

It is most convenient to extract $\xi_3$ from the tadpole relation $N_3=4mn\eps_1^2(1+\rho^2)\xi_3$, which in terms of the integer fluxes
reads $N_3=a_0 b_3 - a_3 b_0$. Combining all the information we readily find
\beq
V_0 = \frac{8 F_V \, m^6 b_3^3 (1+\rho^2)^2}{n^3 N_3^2} \qquad ; \qquad g_s = \frac{m^3 b_3^2 (1+\rho^2)}{F_g \, n^3 N_3} \ .
\label{candata2}
\eeq
Unlike the case when $\rho=0$, in general we cannot keep $m$ and $b_3$ fixed while letting $n \to \infty$. The reason is that the NSNS flux
$b_0$ in (\ref{bzero}) must be an integer. 

The main conclusion is that it is not always possible to obtain small string coupling and cosmological constant. In fact, when $\rho\not=0$, 
there are no vacua with $g_s < 1$ unless the tadpole $N_3$ is sufficiently big. To prove this, notice first that the string coupling
can be rewritten as $g_s=\msm{- b_3 b_0 (1+\rho^2)/(F_s \rho N_3)}$. The most favorable situation occurs when $\rho=-1$ for which $F_s=0.238$.
The smallest allowed NSNS fluxes are $b_0=b_3=2$ (compatible with $\rho=-1$). Hence, the minimum value of the coupling is
$g_s^{min}=\msm{8/(F_s N_3)}$ and $g_s^{min} < 1$ would require $N_3 > 33$. The situation is worse for values of $\rho$ such that multiple vacua can 
appear. The problem is that since such $\rho$'s are rational, $b_3$ must be largish for $b_0$ to be integer.  
Going to a more general gauge does not change the conclusion. 

We have just provided a quantitative, almost analytic, explanation of why there are no perturbative vacua when the flux polynomial
$P_2$ has complex roots and $N_3$ is not large enough. This observation was first made in \cite{stw2} based on a purely numerical analysis.

\end{trivlist}

\section{Final remarks}
\label{sec:end}

In this paper we have investigated supersymmetric flux vacua in a type IIB orientifold with RR, NSNS and non-geometric $Q$-fluxes turned on. 
We enlarged the related analysis of \cite{stw2} by considering the most general fluxes solving the Jacobi identities, and by including variable
numbers of O3/D3 and O7/D7 sources to cancel the flux-induced RR tadpoles. 

Our approach is based on the classification of the subalgebras satisfied by the non-geometric fluxes. A convenient parametrization 
of the $Q$-fluxes leads to an auxiliary complex structure that turns out to be invariant under modular transformations.
Writing the superpotential in terms of this invariant field simplifies solving the F-flat conditions and enables us to obtain 
analytic expressions for the moduli vevs. We have found families of supersymmetric ${\rm AdS}_4$ vacua in all models defined by the inequivalent 
$Q$-subalgebras. General properties of the solutions were discussed in section \ref{sec:lands}.
The vacua typically exist in all cases, provided that arbitrary values of the flux-induced RR tadpoles are allowed.

In type IIB orientifolds with only RR and NSNS fluxes there is a non-trivial induced tadpole that must be cancelled by
O3-planes or wrapped D7-branes. But including non-geometric fluxes can require other types of sources.
For instance, similar to well understood ${\rm AdS}_4$ models in type IIA, the induced flux-tadpoles might vanish implying that sources can be avoided. 
There are also examples in which sources of positive RR charge are sufficient to cancel the tadpoles.
As one might expect, these latter exotic vacua occur in models built using $Q$-fluxes satisfying the non-compact $\mathfrak{so(3,1)}$ subalgebra.
Such solutions might be ruled out once a deeper understanding of non-geometric fluxes has been developed. 

We discussed a simplified set of fluxes but our methods could be used to study other configurations. The starting point
would be the classification of the $Q$-subalgebras consistent with the underlying symmetries.
 
Although our main goal was to explore supersymmetric vacua with moduli stabilized, our results could have further applications.
We have succeeded in connecting properties of the vacua to the underlying gauge algebra and this can help towards extending
the description of non-geometric fluxes beyond the effective action limit. At present one of the most challenging problems in need of new
insights is precisely to formulate string theory on general backgrounds at the microscopic level.

\vspace*{1cm}
                                                                                                                                                         
\noindent                                                                                                                                                         
{\bf \large Acknowledgments}
                                                                                                                                                         
We are grateful to  P.~C\'amara, B.~de Carlos, L.~Ib\'a\~nez, R.~Minasian, G.~Tasinato, S.~Theisen and G.~Weatherill for useful comments.
A.F. thanks the Max-Planck-Institut f\"ur Gravitationsphysik, as well as the Instituto de F\'{\i}sica Te\'orica
UAM/CSIC, for hospitality and support at several stages of this paper, and CDCH-UCV for a research grant No. PI-03-007127-2008. 
A.G. acknowledges the financial support of a FPI (MEC) grant reference BES-2005-8412.
This work has been partially supported by CICYT, Spain, under contract FPA 2007-60252, 
the Comunidad de Madrid through Proyecto HEPHACOS S-0505/ESP-0346, and by the European Union 
through the Marie Curie Research and Training Networks {\it Quest for Unification} (MRTN-CT-2004-503369)  
and {\it UniverseNet} (MRTN-CT-2006-035863).

\newpage

\section*{Appendix: Parametrized RR fluxes}
\label{appA}
\addcontentsline{toc}{section}{\hspace{13pt} Appendix: Parametrized RR fluxes}
\setcounter{equation}{0}
\renewcommand{\theequation}{A.\arabic{equation}}

In this appendix we give the explicit expressions for the original RR fluxes $a_A$ in terms of the 
axionic shifts $(\xi_s, \xi_t)$ and the tadpole parameters $(\xi_3, \xi_7)$ or $(\lambda_2, \lambda_3)$,
depending on the $Q$-subalgebra. For the semidirect sum $\mathfrak{su(2)\oplus u(1)^3}$ and the nilpotent 
algebra there is another auxiliary variable $\lambda_1$ as explained in \ref{ss:rr}.
In all cases there is a non-singular rotation matrix from the $a_A$'s to the new variables.

In principle the $\xi$'s and $\lambda$'s  are just real constants but 
the resulting $a_A$ fluxes must be integers. The exact nature of these parameters can be elucidated starting with the
non-geometric fluxes of each subalgebra. For example, following the discussion at the end of section \ref{subsubso4}, for $\mathfrak{su(2)^2}$
when $\eps_1\eps_2=0$ it transpires that $(\xi_3, \xi_7, \xi_s, \xi_t) \in \mathbb{Q}$.

There is a universal structure in the RR fluxes that is worth noticing. For all $Q$-subalgebras the dependence on the axionic
shift parameters $(\xi_s, \xi_t)$ is of the form
\beqa
a_0 &=& - b_0 \xi_s + 3 c_0 \xi_t \, + \cdots  \nonumber \\
a_1 &=& - b_1 \xi_s - (2c_1 - \tilde c_1) \xi_t \, + \cdots \label{uniRR} \\
a_2 &=& - b_2 \xi_s - (2c_2 - \tilde c_2) \xi_t \, + \cdots \nonumber\\
a_3 &=& - b_3 \xi_s + 3 c_3 \xi_t \, + \cdots \nonumber 
\eeqa
where $\cdots$ stands for extra terms depending on the tadpole parameters.

\begin{trivlist}

\item[{\bf A.1}] \underline{Compact $\mathfrak{su(2)^2}$ background}.
\beqa
\! a_{0} \!\!\! &=& \!\!\! \delta^3(\epsilon_{1} \xi_3 + \epsilon_{2}\xi_s) + \beta^3(\eps_1 \xi_s - \eps_2 \xi_3)  
+ 3 \delta\beta^2(\xi_t - \xi_7) + 3 \beta\delta^2(\xi_t+\xi_7) \nonumber \\[2mm]
\! a_{1} \!\!\! &=&\!\!\! -\gamma\delta^2(\epsilon_{1} \xi_3 + \epsilon_{2} \xi_s)  
- \alpha \beta^2 (\eps_1 \xi_s - \eps_2 \xi_3)  
- \beta(\beta\gamma+2\alpha\delta)(\xi_t - \xi_7) - \delta(\alpha\delta+2\beta\gamma)(\xi_t+\xi_7) \nonumber \\[2mm]
\! a_{2}\!\!\! &=&\!\!\! \delta\gamma^2(\epsilon_{1} \xi_3 + \epsilon_{2} \xi_s)  
+ \beta \alpha^2 (\eps_1 \xi_s - \eps_2 \xi_3)  
+ \alpha(\alpha\delta + 2\beta\gamma)(\xi_t - \xi_7) + \gamma(\beta\gamma +2\alpha\delta)(\xi_t+\xi_7) \nonumber \\[2mm]
\! a_{3}\!\!\! &=& \!\!\! -\gamma^3(\epsilon_{1} \xi_3 + \epsilon_{2}\xi_s) - \alpha^3(\eps_1 \xi_s - \eps_2 \xi_3) 
- 3 \gamma\alpha^2(\xi_t - \xi_7) - 3 \alpha\gamma^2(\xi_t+\xi_7) \nonumber 
\eeqa

\item[{\bf A.2}] \underline{Non-compact $\mathfrak{so(3,1)}$ background}.
\beqa
\! a_{0} \!\! &=& \!\! \delta(\delta^2-3\beta^2)(\epsilon_{1} \xi_3 + \epsilon_{2}\xi_s) 
+ \beta(\beta^2-3\delta^2)(\eps_1 \xi_s - \eps_2 \xi_3)  
- 3(\beta^2 + \delta^2)(\beta\xi_t - \delta\xi_7)\nonumber \\[2mm]
\! a_{1} \!\! &=&\!\! (\gamma\beta^2+ 2\alpha\beta\delta- \gamma\delta^2)(\epsilon_{1} \xi_3 + \epsilon_{2}\xi_s) 
+ (\alpha\delta^2 + 2\beta\gamma\delta - \alpha\beta^2)(\epsilon_{1} \xi_s - \epsilon_{2}\xi_3)
\nonumber \\
&{}& \  + (\beta^2 + \delta^2)(\alpha\xi_t - \gamma\xi_7) + 2 (\alpha\beta + \gamma\delta)(\beta\xi_t - \delta\xi_7)
\nonumber \\[2mm]  
\! a_{2}\!\! &=&\!\! (\delta\gamma^2- 2\alpha\beta\gamma- \delta\alpha^2)(\epsilon_{1} \xi_3 + \epsilon_{2}\xi_s) 
+ (\beta\alpha^2 - 2\alpha\gamma\delta - \beta\gamma^2)(\epsilon_{1} \xi_s - \epsilon_{2}\xi_3)
\nonumber \\
&{}& \   - 2 (\alpha\beta + \gamma\delta)(\alpha\xi_t - \gamma\xi_7) - (\alpha^2 + \gamma^2)(\beta\xi_t - \delta\xi_7)
\nonumber \\[2mm]  
\! a_{3}\!\! &=& \!\!- \gamma(\gamma^2-3\alpha^2)(\epsilon_{1} \xi_3 + \epsilon_{2}\xi_s) 
- \alpha(\alpha^2-3\gamma^2)(\eps_1 \xi_s - \eps_2 \xi_3)  
+ 3(\alpha^2 + \gamma^2)(\alpha\xi_t - \gamma\xi_7)\nonumber
\eeqa

\item[{\bf A.3}] \underline{Direct sum $\mathfrak{su(2)+ u(1)^3}$ background}.
\beqa
a_{0}\!\! &=&\!\! \delta^3(\epsilon_{1} \xi_3 + \epsilon_{2}\xi_s) + \beta^3(\eps_1 \xi_s - \eps_2 \xi_3)  
+ 3 \beta\delta^2 \xi_t - 3\delta\beta^2 \xi_7 \nonumber \\[2mm]
a_{1}\!\! &=&\!\! -\gamma\delta^2(\epsilon_{1} \xi_3 + \epsilon_{2} \xi_s) 
- \alpha \beta^2 (\eps_1 \xi_s - \eps_2 \xi_3)   - \delta(\alpha\delta+2\beta\gamma)\xi_t 
+ \beta(\beta\gamma+2\alpha\delta)\xi_7  \nonumber \\[2mm]
a_{2}\!\! &=& \!\! \delta\gamma^2(\epsilon_{1} \xi_3 + \epsilon_{2} \xi_s) 
+ \beta \alpha^2 (\eps_1 \xi_s - \eps_2 \xi_3)  
+ \gamma(\beta\gamma +2\alpha\delta)\xi_t - \alpha(\alpha\delta + 2\beta\gamma)\xi_7 \nonumber \\[2mm] 
a_{3}\!\! &=&\!\! -\gamma^3(\epsilon_{1} \xi_3 + \epsilon_{2}\xi_s) - \alpha^3(\eps_1 \xi_s - \eps_2 \xi_3) 
- 3\alpha\gamma^2\xi_t + 3 \gamma\alpha^2\xi_7 \nonumber
\eeqa

\item[{\bf A.4}] \underline{Semidirect sum $\mathfrak{su(2) \oplus u(1)^3}$ background}.
\beqa
a_{0}\!\! &=&\!\! \delta^3(\epsilon_2 \xi_s + 3\xi_t) + \beta\delta^2(\epsilon_1 \xi_s - 3\xi_t + 3\lambda_1) 
+ 3 \delta\beta^2 \lambda_2  + \beta^3 \lambda_3 \nonumber \\[2mm]
a_{1}\!\! &=&\!\! -\gamma\delta^2(\epsilon_2 \xi_s + 3\xi_t)
- \msm{\frac13} \delta(\alpha\delta + 2\beta\gamma)(\epsilon_1 \xi_s - 3\xi_t + 3\lambda_1)
- \beta(\beta\gamma+2\alpha\delta)\lambda_2  - \alpha \beta^2\lambda_3  \nonumber \\[2mm]
a_{2}\!\! &=& \!\! \delta\gamma^2(\epsilon_2 \xi_s + 3\xi_t)
+ \msm{\frac13} \gamma(\beta\gamma + 2\alpha\delta)(\epsilon_1 \xi_s - 3\xi_t + 3\lambda_1)
+ \alpha(\alpha\delta+2\beta\gamma)\lambda_2  + \beta \alpha^2\lambda_3  \nonumber \\[2mm]
a_{3}\!\! &=&\!\! -\gamma^3(\epsilon_2 \xi_s + 3\xi_t) - \alpha\gamma^2(\epsilon_1 \xi_s - 3\xi_t + 3\lambda_1)  
- 3 \gamma\alpha^2 \lambda_2  - \alpha^3 \lambda_3 \nonumber 
\eeqa

\item[{\bf A.5}] \underline{Nilpotent $\mathfrak{nil}$ background}.
\beqa
a_{0}\!\! &=&\!\! \delta^3(\epsilon_2 \xi_s + 3\xi_t) + \gamma\delta^2(\epsilon_1 \xi_s + 3\lambda_1) 
+ 3 \delta\gamma^2 \lambda_2  + \gamma^3 \lambda_3 \nonumber \\[2mm]
a_{1}\!\! &=&\!\! -\gamma\delta^2(\epsilon_2 \xi_s + 3\xi_t)
+ \msm{\frac13} \delta(\delta^2-2\gamma^2)(\epsilon_1 \xi_s + 3\lambda_1)
- \gamma(\gamma^2-2\delta^2)\lambda_2  + \delta \gamma^2\lambda_3  \nonumber \\[2mm]
a_{2}\!\! &=& \!\! \delta\gamma^2(\epsilon_2 \xi_s + 3\xi_t)
+ \msm{\frac13} \gamma(\gamma^2 - 2\delta^2)(\epsilon_1 \xi_s + 3\lambda_1)
+ \delta(\delta^2 -2\gamma^2)\lambda_2  + \gamma \delta^2\lambda_3  \nonumber \\[2mm]
a_{3}\!\! &=&\!\! -\gamma^3(\epsilon_2 \xi_s + 3\xi_t) + \delta\gamma^2(\epsilon_1 \xi_s + 3\lambda_1)  
- 3 \gamma\delta^2 \lambda_2  + \delta^3 \lambda_3 \nonumber 
\eeqa

\end{trivlist}

\end{document}